# Need of 6G for the Metaverse Realization

**BARTLOMIEJ SINIARSKI[1], CHAMITHA DE ALWIS[2,7] (Senior Member, IEEE), GOKUL YENDURI[3] (Student Member, IEEE), THIEN HUYNH-THE[6] (Member, IEEE), GÜRKAN GÜR[5] (Senior Member, IEEE), THIPPA REDDY GADEKALLU[3,4] (Senior Member, IEEE) and MADHUSANKA LIYANAGE[1,8] (Senior Member, IEEE)**
[1]School of Computer Science, University College Dublin, Ireland
[2]School of Computer Science and Technology, University of Bedfordshire, United Kingdom
[3]School of Information Technology and Engineering, Vellore Institute of Technology, Vellore, Tamil Nadu, India
[4]Department of Electrical and Computer Engineering, Lebanese American University, Byblos, Lebanon
[5]ZHAW School of Engineering, Switzerland
[6]Department of Computer and Communication Engineering, Ho Chi Minh City University of Technology and Education, Vietnam
[7]Department of Electrical and Electronic Engineering, University of Sri Jayewardenepura, Sri Lanka
[8]Centre for Wireless Communications, University of Oulu, Finland

Corresponding author: Bartlomiej Siniarski (e-mail: bartlomiej.siniarski@ucd.ie).

**ABSTRACT** The concept of the Metaverse aims to bring a fully-fledged extended reality environment to provide next generation applications and services. Development of the Metaverse is backed by many technologies, including, 5G, artificial intelligence, edge computing and extended reality. The advent of 6G is envisaged to mark a significant milestone in the development of the Metaverse, facilitating near-zero-latency, a plethora of new services and upgraded real-world infrastructure. This paper establishes the advantages of providing the Metaverse services over 6G along with an overview of the demanded technical requirements. The paper provides an insight to the concepts of the Metaverse and the envisaged technical capabilities of 6G mobile networks. Then, the technical aspects covering 6G for the development of the Metaverse, ranging from validating digital assets, interoperability, and efficient user interaction in the Metaverse to related security and privacy aspects are elaborated. Subsequently, the role of 6G technologies towards enabling the Metaverse, including artificial intelligence, blockchain, open radio access networks, edge computing, cloudification and internet of everything. The paper also presents 6G integration challenges and outlines ongoing projects towards developing the Metaverse technologies to facilitate the Metaverse applications and services.

## I. INTRODUCTION

The term 'Metaverse' has been coined to further facilitate the digital transformation in every aspect of our physical lives [1]. The Metaverse is a virtual world where you can live a synchronous life through your avatar. The concept is similar to an online game, however, instead of shooting targets or driving cars, users will be engaged in real-life activities. These activities could include attending meetings, catching up with friends, attending music festivals, going door to door selling digital collectables, or buying and selling land, apartments or assets. Virtual interactive worlds or early Metaverses have already been introduced primarily in video games with releases such as Fortnite, Minecraft, Decentraland, Ifland. The list isn't extensive and users are gravitating toward other Metaverse ecosystems that are emerging today. The Metaverse embraces a social interaction accelerated through a virtual environment and driven by novel technologies such as Web 3.0, 5G, Artificial Intelligence (AI) and Extended Reality (XR). The XR - which includes everything from Virtual Reality (VR) to Mixed Reality (MR) to Augmented Reality (AR) and haptics - have enormous potential to transform both industry and society. The widespread adoption of XR was slowed down recently by a number of issues including limited processing power, storage and battery life of small head-mounted displays (HMDs). The 5G made it possible to overcome some of these challenges by offloading a portion of XR processing to the mobile network edge. In addition to this, the 5G QoS framework makes it possible to establish QoS flows that provide optimized network treatment for specific traffic flows, in addition to the default QoS flow used for mobile broadband (MBB). Such additional QoS flows can be established either using 5GC QoS-exposure application programming interfaces to communicate service requirements or by traffic detection together with pre-provisioned service requirements, such as relying on standardized 5G QoS identifier characteristics.

Although the Metaverses have the potential to be transformational for both business and society, widespread adoption has previously been hindered by issues such as heat generation and the limited processing power, storage, and battery life of small form factor head-mounted devices. The time-critical communication capabilities in 5G make it possible to overcome only some of these challenges by offloading XR processing to the mobile network edge. By evolving the already existing 5G or B5G networks, mobile network operators are in an excellent position to enable the realization of the Metaverse on a large scale. The 6G aims to achieve high spectrum and energy efficiency, low latency, and massive connection due to the exponential growth of the Internet of Things (IoT) devices. 6G will also effectively link the physical, and digital worlds by providing seamless and ubiquitous services such as extreme-scale environmental monitoring and control, virtual reality/virtual navigation, telemedicine, digital sensing, and robotics. This will result in a network that connects us to one another, to information, to knowledge, and to purpose. As a result, 6G networks will enhance the efficiency of technologies such as computer vision, blockchain, AI, the IoT, robotics, and user interfaces which are critical for the metaverse realization. In summary, 6G will enhance every feature of the 5G network that benefits the user to improve areas such as smart cities, farming, manufacturing, and robots. 6G will provide enhanced productivity, capabilities, and better user experiences. The main use of 6G in the Metaverse is summarized below:

**Near-zero-latency**: In virtual interaction, 6G will continuously provide users with a near-zero-latency sensory interconnection experience, such as the user's virtual movement in the Metaverse, virtual meetings, virtual paintings, and other interactive, immersive holographic experiences.

**New services**: 6G is the main driver of multiple new service models. For example, 6G communication technology provides users with precise service models in autonomous driving, industrial control, e-health, Internet robots, and autonomous systems, bringing a more convenient lifestyle.

**Upgraded real-world infrastructure available for use in the Metaverses**: 6G infrastructure mainly includes information infrastructure, fusion infrastructure, and innovation infrastructure. In particular, the 6G communication system integrates infrastructure such as ground, UAV, and satellite Internet. 6G also features high bandwidth, low latency, strong reliability, and global coverage.

### A. MOTIVATION

The main motivation of this paper is to realize if mobile network operators can enable large-scale XR and at the same time further development of Metaverses by introducing time-critical communication capabilities in 6G networks. The 5G networks already contribute to considerable improvement in data rate, latency, and packet loss since the last network generation (4G) and users already enjoy comfortable viewing experiences. However, as the resolution of video increases from 4K to 8K and 12K/24K for 3D video and the number of worldwide users increases, the 5G network will not be sufficient to support many use cases. Some of the main cloud and network providers are defining the evolution of the service experience into the fair-experience, comfortable experience, and ideal-experience phases [2], where each has its own network KPI requirement to be met. Table 1 summarizes those KPI requirements based on different use cases envisaged to be a part of future metaverses. In this work, we aim to establish and explain the main advantages of providing the Metaverse services over 6G and provide an overview of the technical requirements. Furthermore, we aim to establish what role will 6G play in the Metaverse operation and if the envisaged architecture of 6G will be capable of supporting the upcoming technology portrayed by the tech industry.

### B. RELATED SURVEYS AND CONTRIBUTIONS

Our work is exclusively focused on the networking aspects of the Metaverse and the role that 6G will play in the Metaverse deployment. Though there are some Metaverse-focused surveys we found it is lacking a comprehensive, and detailed discussion on the role of B5G/6G technologies as indicated by Table 2. The table also includes the limitations of the related works in the context of technical challenges, security and privacy, and research directions, which we have already addressed in this paper. The surveys [3] and [4] investigate technical aspects and security threats comprehensively. However, those papers are not focused on the future networks and the role of 6G in the Metaverse specifically. The surveys [5], [1] and [6] include an interesting view on the potential use of the Metaverse in different domains and clearly define network requirements. The limitations in [5], [1] and [6] include the lack of coverage of future network aspects and the discussion on the security and privacy issues is weak. Surveys [7] and [8] discuss implementation challenges and analyze the fusion of 6G-enabled edge with the Metaverse, however, the security issues and research directions are only partially covered.

Therefore, we contribute to addressing this gap in our work on the comprehensive discussion on 6G for the Metaverse.

### C. PAPER ORGANIZATION

The rest of this paper is organized as follows. Introduction and discussion of the role of 6G networks in the Metaverse are presented in Section I. Section II covers the expected improvements from 5G to 6G and the impact it will have on the Metaverses. Section III investigates the state-of-the-art solutions provided by 6G for the Metaverse from technical perspective, followed by Section IV that discusses in detail how different 6G technologies will help to achieve the Metaverse aims. Section V identifies expected 6G challenges that would have to be approached before the introduction of Metaverses to wider community. Finally, Section VI provides an overview of related research projects.

TABLE 1. Network KPI requirements in different phases of cloud/ the Metaverse development

| Type of interaction / use case | Network KPI requirement | Fair-experience<br>In the fair-experience phase, most content is 4K, and the terminal screen resolution is 2K to 4K. | Comfrotable-experience<br>In the comfortable-experience phase, most content is 8K, and the terminal screen resolution is 4K to 8K | Ideal-experience<br>In the ideal-experience phase, most content is 12K or 24K. The terminal screen resolution is 8K to 16K. |
|---|---|---|---|---|
| **Weak-interaction**<br>Users select view and location, but do not interact with entities in the virtual environment. For example IMAX, 360 video, live broadcast, music, education. | Bitrate | ≥ 40 Mbit/s (4K) | Full-view: ≥ 90 Mbit/s<br>FOV: ≥ 50 Mbit/s | Full-view: ≥ 290 Mbit/s (12K)<br>≥ 1090 Mbit/s (24K)<br>FOV: ≥ 155 Mbit/s (12K)<br>≥ 580 Mbit/s (24K) |
| | Bandwidth requirement | ≥ 60 Mbit/s (4K) | Full-view: ≥ 140 Mbit/s<br>FOV: ≥ 75 Mbit/s | Full-view: ≥ 440 Mbit/s (12K)<br>≥ 1600 Mbit/s (24K)<br>FOV: ≥ 230 Mbit/s (12K)<br>≥ 870 Mbit/s (24K) |
| | Recommended network RTT | ≤ 20ms | ≤ 20ms | ≤ 20ms |
| | Packet loss requirement | ≤ 9e-5 | ≤ 1.7e-5 | ≤ 1.7e-6 |
| **Strong-interaction**<br>Users can interact with virtual envirnomnets through interactive devices. The virtual space displayed needs to respond to interactions in real time. For example gaming, fitness, social networking, real estate, engineering, healthcare, shopping. | Bitrate | ≥ 40 Mbit/s | ≥ 90 Mbit/s | ≥ 360 Mbit/s (8K)<br>≥ 440 Mbit/s (16K) |
| | Bandwidth requirement | ≥ 80 Mbit/s | ≥ 260 Mbit/s | ≥ 1000 Mbit/s (8K)<br>≥ 1500 Mbit/s (16K) |
| | Recommended network RTT | ≤ 20 ms | ≤ 15 ms | ≤ 8 ms |
| | Packet loss requirement | ≤ 1e-5 | ≤ 1e-5 | ≤ 1e-6 |

## II. 6G AND THE METAVERSE: PRELIMINARIES

The preliminary introduction to 6G and the Metaverse is presented in this section, followed by the role of 6G in the Metaverse.

### A. PRELIMINARY TO 6G

Since the middle of 2019, commercial 5G mobile networks have been standardized and deployed globally, with significant coverage in some countries. Numerous new applications and use cases are being developed, placing existing networks' capabilities to the test. The capacity of current 5G networks to handle the Internet of Everything (IoE), holographic telepresence, collaborative robotics, and deep-sea and space tourism is limited [9]. This has prompted researchers to reconsider and work toward the development of the next generation of mobile communications networks called the sixth-generation of mobile networks 6G. Each time mobile communication technology is upgraded and iterated, its performance metrics improve by a factor of ten to hundred times over the preceding generation [10]. Researchers from all over the world propose AI/machine learning (ML), quantum communication/quantum machine learning (QML), blockchain, tera-hertz and millimetre wave communication, tactile Internet, non-orthogonal multiple access (NOMA), small cell communication, fog/edge computing, etc. as the key technologies for the realisation of 6G communications. 6G aims to achieve high spectrum and energy efficiency, low latency, and massive connection due to the exponential growth of the IoT devices. 6G will make feasible intelligent traffic, environmental monitoring and control, virtual reality/virtual navigation, telemedicine, digital sensing, high definition (HD), and full HD video transmission in connected drones and robotics. 6G will also effectively link the physical, and digital worlds. This will result in a network that connects us to one another, to information, to knowledge, and to purpose. 6G wireless networks operate in the terahertz band, with a peak rate of 1T b/s and a network with ultra-reliable and low-latency communication (URLLC) of less than 1 ms, considerably improving the overall quality of experience (QoE) for consumers [11]. 6G has a high positioning accuracy of 1 m outdoors and 10 cm indoors [12] which also improves positioning accuracy of deep-sea and space tourism. 6G utilises endogenous security technology to increase its resistance to unknown security threats [13]. As a result, 6G networks can enhance the efficiency of technologies such as computer vision, blockchain, AI, the IoT, robotics, and user interfaces [14]. The architecture of 6G is depicted in Fig. 1

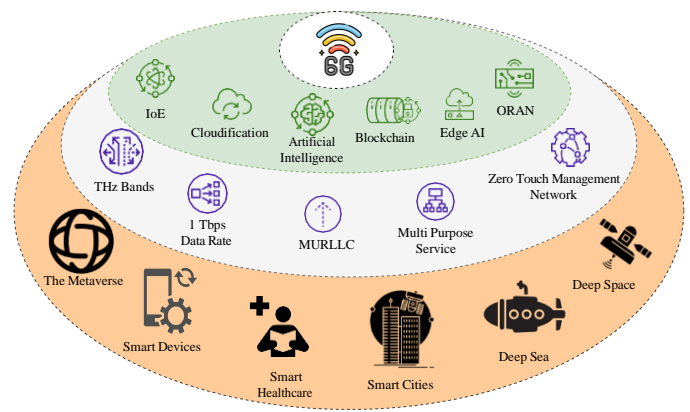

FIGURE 1. 6G Architecture

To summarize, 6G will enhance every feature of the 5G network that benefits the user. 6G will improve areas such as smart cities, farming, manufacturing, and robots. 6G will provide enhanced productivity, capabilities, and better user experiences. Improved and expanded functionality is an in-

TABLE 2. Summary of Important Surveys on 6G and its role in the Metaverse

| Ref | Technical aspects and challenges | Security and privacy issues | The role of 6G in Metaverse | Research directions (6G) | Remarks | Limitations |
|---|---|---|---|---|---|---|
| [5] | H | M | H | M | This paper aims to show the roadmap to the Metaverse in terms of communication and networking in 6G, including requirements (limited) and challenges for 6G to realize the Metaverse, and discussing the fundamental technologies to be integrated in 6G to drive the implementation of the Metaverse. | The paper is missing some important references and the requirements are not discussed in detail except for what is depicted in Fig. 2. |
| [3] | H | M | L | L | The paper investigates AI-based methods concerning six technical aspects that have potentials for the Metaverse: natural language processing, machine vision, blockchain, networking, digital twin, and neural interface, and being potential for the Metaverse. | The discussion on attacks on AI in the Metaverse is missing, subsequently the paper should discuss the role of AI in future networks from the security and privacy perspective. |
| [1] | H | M | L | M | The technology enablers are discussed in details including latest state-of-the-art tools. The survey paper includes an interesting discussion on user-centric factors. | The paper does not cover any aspects of future networks and how those could play an important role in creating Metaverses. |
| [4] | M | H | L | L | The security threats to the Metaverse are explained comprehensively. The paper includes the countermeasures to those threats. From the security and privacy perspective, it is a comprehensive survey. | The future networks enablers are not discussed in this paper. The paper provides good state-of-the-art survey, however it is lacking future directions especially in the networking aspect. |
| [6] | M | L | L | M | The paper provides an interesting view on the potential use of the Metaverse in medical domain including a proposed process of patient treatment using the Metaverse technology. | This paper does not cover any security or privacy issues or the importance of future networks in designing meta worlds. |
| [7] | H | M | L | M | This survey discusses how enablers of the Metaverse can be implemented at a scale at mobile edge networks. The implementation challenges are also discussed in this survey. | The survey mentions the role of B5G/6G, but it doesn't cover how 6G will enable the Metaverse. |
| [8] | M | L | M | M | This survey analyzes the fusion of 6G-enabled edge AI and the Metaverse, and introduces the features, architecture, technologies, and applications of the Metaverse. | The paper only partially covers privacy and security issues. The 6G requirements are discussed only to certain level and the 6G enablers are not covered in full. |
| Our survey | H | M | H | H | In this paper we focus on 6G technology and how it enables the deployment of Metaverses. We focus heavily on specific requirements and cover each in detail as supposed to provide the reader with general overview. The security and privacy challenges are covered in depth despite of this not being the main focus of this paper. We clearly identify current research projects and future research directions, which is something missing in most survey papers that were investigated. | There is still space to discuss other aspects once meteverses are explored in more depth, in particular social aspects such as fairness, social acceptance, accountability, and community ownership. |

Low Coverage: The paper did not consider this area or only very briefly discussed it through mentioning it in passing
Medium Coverage: The paper partially considers this area (leaves out vital aspects or discusses it in relation to other areas without a specific focus on it)
High Coverage: The paper considers this area in reasonable or high detail evitability over successive generations. Even with 6G, this will be the case. 6G will improve upon 5G by optimising and lowering costs to increase adoption. Information management and consumption will be simplified with the advent of 6G's new human-machine interfaces. The touchscreen interface of the future will instead be controlled by voice instructions, gestures, or even brain signals. The comparison of the features related to 5G and 6G are depicted in Table 3

### B. PRELIMINARY TO THE METAVERSE

The Metaverse is a network of three-dimensional virtual environments dedicated to social interaction. It is frequently depicted in futurism and science fiction films. The worldwide virtual environment will be made possible by the usage of VR and AR devices [15]. The term "Metaverse" is not entirely unfamiliar in the technological world. Neal Stephenson coined the term "Metaverse" in 1992. His science fiction novel Snow Crash envisioned an online universe in which people may explore and escape the physical world using digital avatars [16]. Decades later, major technology firms like Meta, Axie Infinity, The Sandbox, and Decentraland have begun developing their versions of a futuristic Metaverse. The overview of the enabling technologies, services, and technical requirement is depicted in the Fig. 2

**TABLE 3.** The comparison of 5G and 6G Features

| Features | 5G | 6G |
|---|---|---|
| Data Rate | 1 Gbps to 20 Gbps. | 1 Tbps. |
| Application Types | Enhanced Mobile Broadband. Ultra-Reliable Low Latency Communications. Massive Machine Type Communications. | Massive Broadband Reliable Low. Latency Communication. Massive-URLLC. Human-Centric Services. Multi-Purpose Services. |
| Device Types | Smartphones, Drones, and Sensors. | Sensors & DLT devices, BCI and XR equipment, CRAS, and Smart implants. |
| Frequency Band | Sub 6 GHz and mm wave for fixed access. | Sub 6 GHz mm wave for mobile access exploration of THz bands. Non-RF bands. |
| Latency | 5 ms | <1 ms |
| Architecture | Dense sub 6 GHz smaller BSs with umbrella macro BSs. Mmwave small cells of about 100 meters. Cell free smart surfaces at high frequencies. | Cell free smart surfaces at high frequencies. Temporary hotspots provided by drone-mounted BSs. Trials of tiny THz cells. |
| Spectral and Energy Efficiency Gain | 10 x in $bps/Hz/m^2$. | 1000 x in $bps/Hz/m^2$. |
| Traffic Capacity | 10 $Mbps/m^2$. | 1 to 10 $Gbps/m^2$. |
| Reliability | $10^{-5}$. | $10^{-9}$. |
| Localization Precision | 10cm. | 1cm in 3D. |
| User Experience | 50 Mbps. | 10 Gbps. |
| Mobility | 500 km/h | 1000 km/h |
| Connection density | $10^6$ devices/$km^2$ | $10^7$ devices/$km^2$ |

1) Enabling Technologies of the Metaverse

The immersive experience of the Metaverse will be enabled by cutting-edge technologies such as blockchain, AR and XR, AI, and the IoT.

- Blockchain: Blockchain technology enables decentralised and transparent digital proofs of ownership, collectibility, value transfer, governance, accessibility, and interoperability in the Metaverse [17]. Blockchain also enables individuals to exchange assets while working and interacting in the Metaverse.
- Extended reality: XR enables the creation of 3D computer-rendered virtual environments in the Metaverse. XR allows users to interact with these virtual goods through head tracking devices or physical controls [18]. As XR technology matures, it will be able to broaden the Metaverse experience by including physical simulations using XR equipment. Users will then have the ability to sense, hear, and interact with people from all around the world.
- Artificial intelligence: AI will allow users of the Metaverse to construct incredibly realistic avatars and have multilingual accessibility. AI will help people make better decisions in the Metaverse. A better human-computer interface will also be provided by AI [19]. AI can also help detect, prevent, and recover from cyber attacks in the Metaverse.
- Internet of things: IoT will enable the Metaverse to map data from real life and emerge it into virtual reality [20]. The data provided from the IoT devices to the Metaverse will help professionals to solve real-world problems.

The Metaverse with the help of IoT will support the users to collect real-time data-driven decisions with minimum need for training and computation.
- Edge Computing: Edge computing enables mobility and the border-less Metaverse for the users [21]. Edge computing improves the availability of data in the Metaverse by bringing data closer to end consumers for retrieving and storing data at remote data centre locations. Edge computing will help data transferring with ultra-reduced latency in the Metaverse which will help the users to make quick and effective decisions.

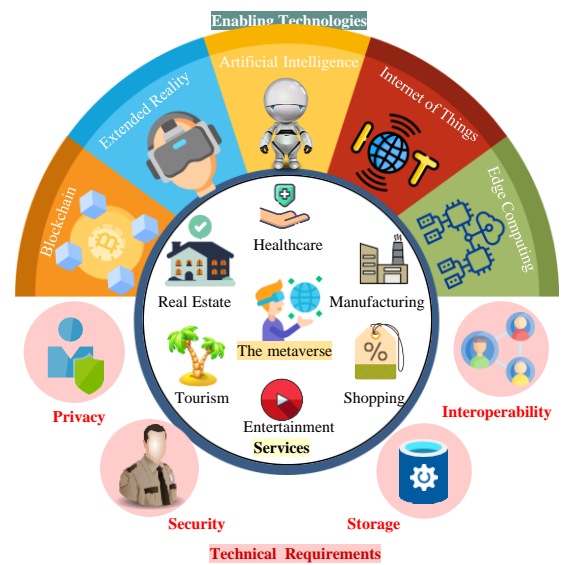

**FIGURE 2.** Preliminary to the Metaverse

2) Applications of the Metaverse

The Metaverse has made its way into many sectors, capturing the enthusiasm of entrepreneurs across the world. The Metaverse will have a huge impact on applications like healthcare, real estate, manufacturing, tourism, Entertainment, and shopping.
- Healthcare: Smart healthcare has contributed to resolving several healthcare difficulties, including linking patients to doctors situated throughout the world during the COVID-19 epidemic. This prepared the door for the application of the Metaverse in healthcare, which is facilitated by medical IoT, VR, and AR [22]. The Metaverse gives users control over how the virtual and physical worlds interact. This enhances doctors' ability to provide consistent and customised patient care. Through the use of VR technology, the Metaverse can aid in remote health monitoring, clinical data collection, and improved robotic surgery, while 3D immersive technology will enable surgeons to practise through simulations that will raise their success rate in the real world.
- Real Estate: The Metaverse allows organisations to establish retail and experience centres on its virtual land [23]. Rather than downloading many applications, users can access the Metaverse, which contains all currently available applications. As a result, the value of virtual land will increase. Property ownership in the Metaverse is limitless, and owners are free to use their virtual holdings. Digital property owners can make, run, lease, and build billboards for advertising.
- Manufacturing: Manufacturers can create digital factories in the Metaverse to assist in organising production and effective usage of machinery. This allows simulation of human-machine interaction throughout the manufacturing process [24]. As a result, firms can use virtual production systems to train new employees and staff on how to use them in the real world, which would boost the manufacturing of products with a very low error rate. The metaverse also allows mass personalization of the product and allows the user to track the product from its development to delivery, which will improve the trust of the users in the organization.
- Tourism: The Metaverse has the potential to create the most immersive experiences ever seen in the tourism sector. The Metaverse allows hotel chains, tourism boards, online travel agencies, and other businesses to advertise their services [25]. Users can virtually visit those locations and decide whether or not to visit them in person. They can go through two distinct locations without leaving their homes, comparing and evaluating places through the use of 3D imagery. The Metaverse will give users an experience that will be better than any kind of communication that exists in the present day, including video and audio interaction.
- Entertainment: The Metaverse will completely revolutionise entertainment with its rich 3D environment and avatars. Entertainment's growth is highly linked to the development of VR in the Metaverse. The Metaverse-based entertainment, including movies and video games, can be enjoyed in a virtual world that users can access from the comfort and privacy of their home [26]. It also allows users to attend virtual concerts and sporting events from first-row seats or to ride virtual roller coasters at theme parks.
- Shopping: Customer experiences in the Metaverse will evolve constantly as a result of XR technologies, and organisations selling products in metamalls will have more creative freedom to express themselves and attract customers than they do in traditional shopping. These spaces will encompass much more than the basic services seen on the majority of e-commerce sites today, including user engagement, avatar customization, event attendance, and skill acquisition [27]. Furthermore, the products sold in the Metaverse will include both physical and virtual items. Consumers may feel and touch the object with the use of sensors, which will completely alter the traditional buying experience. Additionally, customers can purchase things on the go while engaged in real-world activities.

3) Technical Requirements of the Metaverse

Privacy, security, storage, and interoperability are the important technical requirements of the Metaverse.
- Privacy: The Metaverse is a social platform that employs interactive technology such as VR, AR, and AI that requires sensitive user data. Since behavioral-learning and recommender systems collect vast quantities of personal information, they pose a threat to the privacy of the Metaverse users [28]. Therefore, the use of such technologies poses a substantial risk to the privacy of users' data. The Metaverse must guarantee privacy protection for such sensitive information, and users must have complete control over their data, which will increase their trust in the Metaverse. Even though blockchain technology can help protect privacy in the Metaverse, there are no specific rules designed for privacy protection in the Metaverse, which makes it a critical requirement.
- Security: In the Metaverse, attackers and AI bots can and will emerge from any location and at any time. The Metaverse networks should have a high level of security, and related protocols to incorporate continuous awareness into these networks [5]. In addition to existing passwords, multi-factor authentication, enhanced firewalls, and advanced threat detection technologies, the Metaverse must be incorporated with higher transparency and analysis to detect anomalies and uncover malicious activities to maintain user security. Data must be secured and safeguarded during transmission and storage. To assure the security of the Metaverse in the future, it is vital to draw on and build upon information from the past.
- Storage: The Metaverse is a collection of technologies. It is a huge concept which involves the simultaneous integration of multiple technologies. The list includes high-performance networks, sophisticated computing and sensors, hardware, AR/VR, AI, 3D modelling, blockchain, IoT, and many other technologies. The data produced from these technologies and their related application will be enormous [29]. The formation of the Metaverse itself necessitates voluminous data storage. Decentralised storage based on blockchain technology can be used to store this massive amount of data. This storage distributes data to numerous independent network nodes using open-source applications and algorithms. It also improves the privacy of data, the redundancy of data backups, and the transparency of data in the Metaverse.
- Interoperability: Interoperability across services, technology, and virtual worlds is a crucial aspect of the

Metaverse [30]. A cross-chain protocol is an optimal approach for maintaining interoperability between diverse Metaverse services, technologies, and configurations. Among other protocols, this one permits the exchange of assets like avatars, non-fungible tokens, and currency. To make the Metaverse more interoperable, different devices that use the same technology need to follow the same rules and standards.

### III. 6G FOR THE METAVERSE: TECHNICAL PERSPECTIVE

This section investigates the state-of-the-art solutions provided by 6G for the Metaverse from the technical perspectives, including validation of digital assets, cross-platform integration, efficient support of AI, privacy and security, high-speed data connection, content creation and storage, user interactivity, low latency communication, and computer vision.

#### A. 6G FOR VALIDATION OF DIGITAL ASSETS IN THE METAVERSE

Non-fungible Token (NFT) is one of the key components of a digital asset in the Metaverse. A visual asset, such as a virtual building, can be represented by an NFT that can be represented as a digital asset with unique data coding. When a purchaser buys an NFT, a private digital key password will be generated, that can certify that the purchaser owns a particular digital asset. Through this private key, the owner of the NFT can sell or transfer the digital asset to others [31]. The blockchain associated with the specific Metaverse will record the NFT for a digital asset in the Metaverse. For instance, the Ethereum blockchain stores "Decentraland Metaverse", which is highly popular. Ethereum records the digital asset transactions of the NFT for Decentraland [32]. Digital assets in the Metaverse can be created in the form of NFTs by the users. These digital assets can be anything ranging from virtual goods to digital art to virtual real estate, which is minted into the NFTs that are securely stored on the blockchain. The owners of these digital assets can see these digital assets which are in the form of NFTs in the form of crypto for purchasing other digital assets in the Metaverse [33]. Content creators and artists can afford to have an opportunity to monetize their assets uniquely due to NFTs and blockchain. For instance, artists need not depend on auction houses or galleries for selling their art. The artists can sell their art directly to the customer in the form of an NFT that allows them to keep the majority of the profits. The artists can even receive royalties whenever their art is sold to a new customer. Like art, the other digital assets also can be traded through NFTs in the Metaverse [34]. The process of creating NFTs and transferring them from one virtual world to another requires a network that is highly reliable and secure.

Digital assets in the Metaverse, represented by NFTs are verified and tracked on the blockchain. Every NFT will have a unique transaction hash that makes it non-replicable. All the transactional data related to the NFTs are collected by the blockchain and are stored in blocks, that forms a blockchain. The information stored in the blockchain is stored forever and can be viewed and verified by the public. Verification of the digital assets in the Metaverse that has AR and other MR technologies incorporated, needs significant amount of bandwidth to create a more immersive experience add also to reduce the load times. The validation and verification of the digital assets in the blockchain incurs heavy computation in the blockchain, which needs significant bandwidth so that the users can see the results in near real-time, as depicted in Fig. 4 . The transactions between the different entities in the Metaverse are also powered by the consensus mechanism of the blockchain, which requires huge amounts of data transfer between nodes. This creates a requirement for a network that is both transparent and capable of real-time communication. These challenges faced during the creation, transfer, and validation of digital assets in the Metaverse can be solved by 6G due to its low latency, reliability, transparency, and high-speed connectivity [35].

#### B. 6G FOR CROSS PLATFORM INTEGRATION/INTEROPERABILITY IN THE METAVERSE

One of the hurdles in realizing the full potential of the Metaverse is the lack of interoperability. Lack of interoperability [36], [37] is a major hurdle in a mass adaption of the Metaverse that makes the navigation of the users free from one Metaverse application to the other challenging. The Metaverse should mimic the interoperability that is experienced in the physical world. For instance, in the real/physical world, we can take physical assets/objects from one place to another easily. The users in the Metaverse too should be able to navigate seamlessly and freely to other Metaverses. This is possible through interoperability that can form a global interconnected Metaverse where various Metaverses are integrated across the platforms as experienced in the real world [38].

Realization of interoperability in the Metaverse is a significant challenge as heavy objects such as digital avatars, 3D holograms etc. have to be navigated across in feature-rich Metaverse in near real-time. It requires a communication infrastructure with high bandwidth and low latency. 6G network with its high bandwidth and ultra-reliable low latency communication infrastructure can solve the issue of seamless communication in the Metaverse, as depicted in Fig. 5 . With the help of supporting technologies like ORAN and ZSM, the 6G network can be the common platform that provides an interoperable infrastructure for multiple Metaverses. Network slicing, software-defined networking, symbiotic radio, and network function virtualization are the 6G techniques that promote network interoperability and agility in the Metaverse. Intelligent collaboration between diverse wireless signals is supported by symbiotic radio. The SDN/NFV offers open interfaces that facilitate interoperability between several Metaverses and assist produce network slices for any vertical application such as gaming and shopping over the common

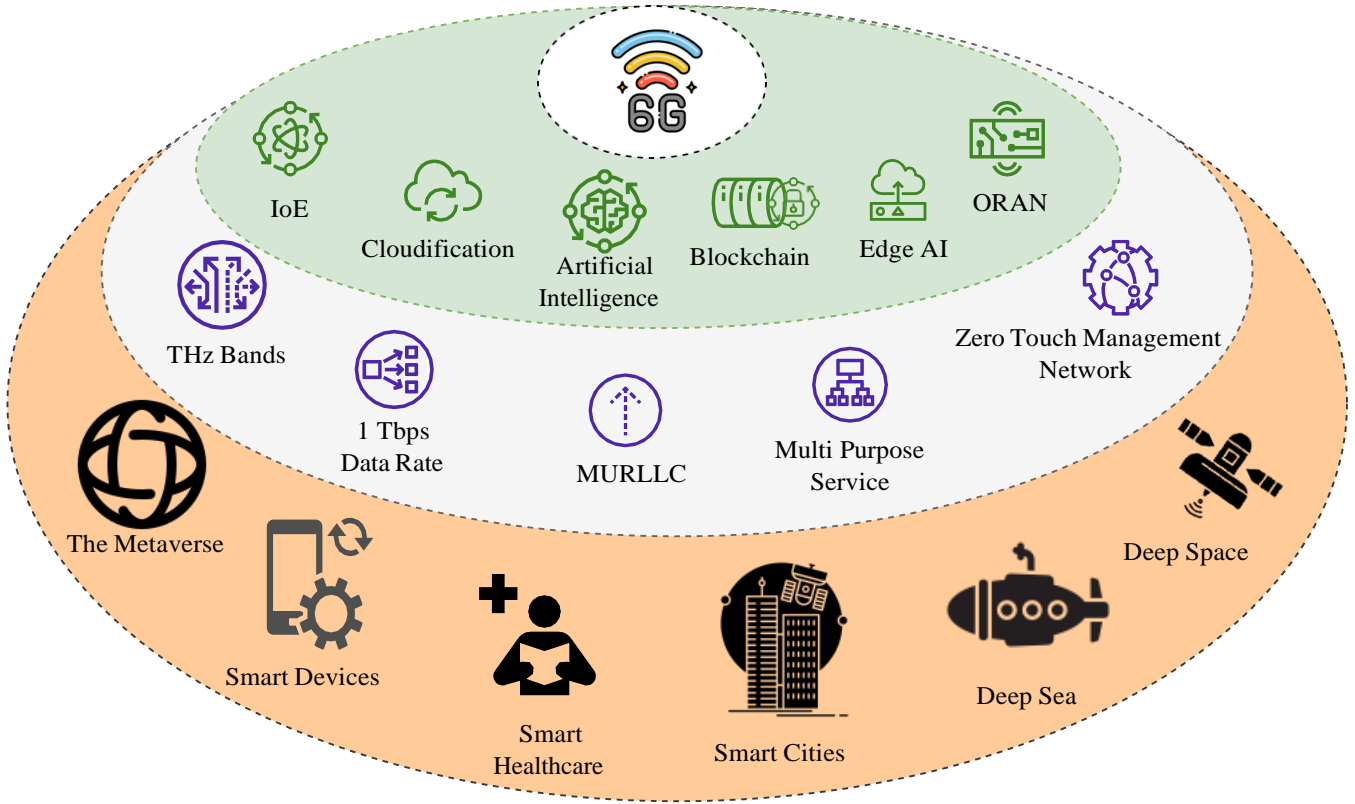

**FIGURE 3.** 6G for the Metaverse:Technical Perspective

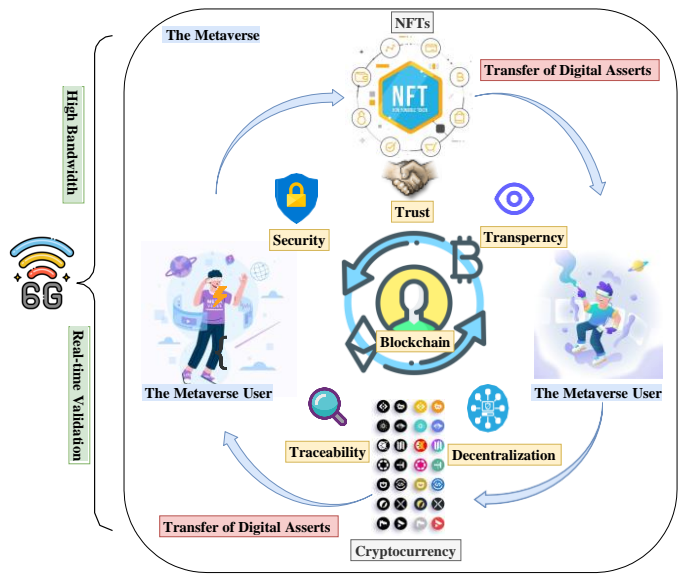

**FIGURE 4.** 6G for Validation of Digital Assets in the Metaverse

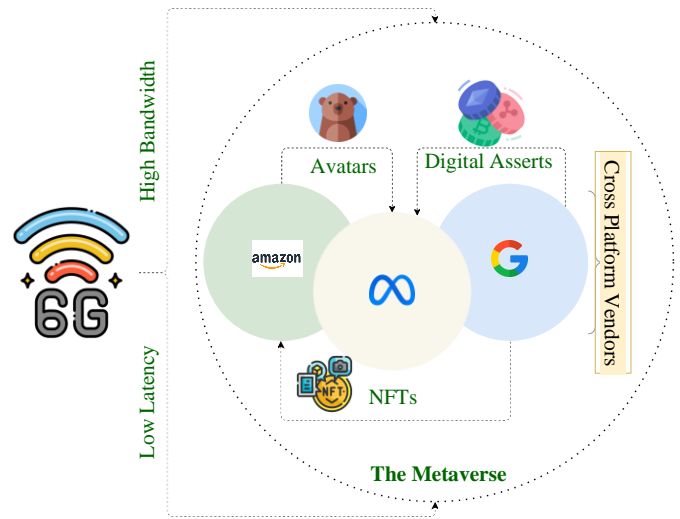

**FIGURE 5.** 6G for Cross Platform Integration/Interoperability in the Metaverse

physical infrastructure among different Metaverses [39].

### C. 6G FOR EFFICIENT SUPPORT OF AI IN THE METAVERSE

The Metaverse is a virtual world where the users will play games, interact with each other and the 3D objects in the virtual world, and build things in the virtual world. VR and AR along with blockchain and AI are the key enabling technologies in realizing the Metaverse. The applications of AI in the Metaverse include speech processing, content analysis, computer vision, etc [3]. These applications of AI can be used to help build important components of the Metaverse as discussed below:

**Avatars:** Avatars is one of the important and interesting concepts of the Metaverse, where people in the physical world will create a digital avatar in the virtual world. People would like to get creative in the virtual world and they would like to see themselves in a different way which may not be possible in the physical world. They can change their clothing, hairstyle, and body language, which is not their regular norm in the real world. AI plays a major role in the users designing their avatars in the virtual world. AI can be used to analyze 3D scans or user images to create accurate, realistic, and innovative avatars [40]. Some organizations such as Ready Player Me are making use of AI to create avatars for the Metaverse.

**Digital Humans:** In the Metaverse, 3D chatbots are termed as digital humans. The digital humans respond and react to the actions of humans in the virtual world. They are usually non-playing characters that can be a character in a game of virtual reality whose actions and responses depend on a set of rules or automated scripts. They try to understand what the users are communicating by listening and observing them. Human-like interactions and conversations can happen in the Metaverse between humans and digital humans through body language and speech recognition [41]. AI plays a significant role in the successful implementation of digital humans in the Metaverse. Some of the key functionalities of digital humans like speech recognition, body language identification, object detection, etc. can be realized through AI [6].

**Language Processing:** In the Metaverse users from across the globe can communicate and interact easily without language barriers. This is made possible with the help of AI. AI can break the human language such as English into a format that can be read by machines. The AI can then analyze the sentences, and respond to the users in their language [42].

From the above discussion, it is obvious that AI plays a significant role in the realization of some of the key features of the Metaverse. In the Metaverse, huge volumes of heterogeneous big data will be generated at a very fast rate. 6G, with its characteristics such as fast communication infrastructure, and near real-time processing, can help in processing/analyzing this big data to uncover the patterns existing in the data that trains the AI/ML algorithms in near real-time to make quick decisions/predictions through which several components of the Metaverse can communicate easily.

### D. 6G FOR HIGH SPEED DATA CONNECTION IN THE METAVERSE

The wide adaption of AR and VR technologies is the key to the transition to the Metaverse. It is expected that data usage to be increased by 20 times to what is being used today due to the revolution of the Metaverse by 2022. To realize the full potential of the Metaverse with real-time experience of AR and VR technologies, truly immersive 3D experiences. The end-users should be able to access high-speed data connections that can deliver the data at speeds of approximately 1 Gbps [43]. Some of the key requirements that will be needed to realize the true potential of the Metaverse are as follows:

- To create virtual reality worlds in real-time, high-speed data connection is required.
- The communication infrastructure should high-speed transmission in near real-time with very low latency, typically, below 10 milliseconds.
- The existing 4K video resolution may not be sufficient to convey the pixels for creating immersive worlds. Higher-resolution videos have to be supported by the data carriers.
- Next-generation video compression techniques that can compress and decompress huge data files in the Metaverse in real time are the need of the hour.

The key features of 6G with high bandwidth and URLLC [44] promise is a key enabling technology to realize the high bandwidth requirement of the Metaverse. The use of Edge AI-enabled 6G can also help applications and the Metaverse devices address these issues. Edge AI is the combination of edge computing and AI to run machine learning tasks directly on connected edge devices. Edge AI computes and processes data locally, which helps the Metaverse devices be efficient and responsive in their communication. This also reduces the amount of data sent from the Metaverse devices to the cloud, thereby saving a huge amount of bandwidth.

### E. 6G FOR EFFICIENT USER INTERACTION IN THE METAVERSE

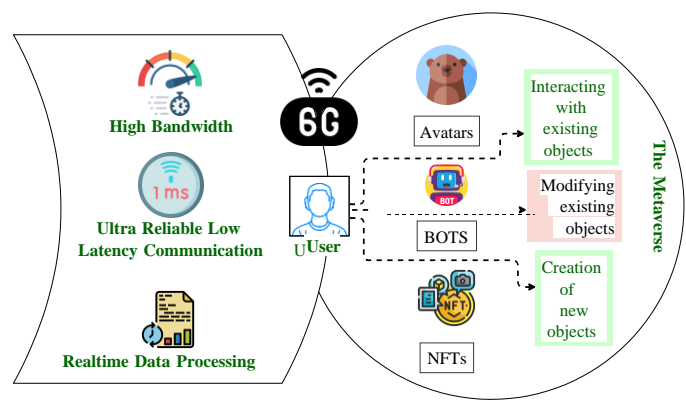

**FIGURE 6.** 6G for Efficient User Interaction in the Metaverse

The Metaverse enables the interaction between real-world entities and virtual objects. It is a digital environment that incorporates social networking, real estate, online gaming, AR, VR, and cryptocurrencies. In the Metaverse, with the help of virtual objects, sounds, and other sensory input, AR tries to enhance the user's perception of the real world [5]. Each time a user enters the Metaverse, the objects around them undergo a dynamic transformation based on the requirements.

Everything in the Metaverse is constantly changing, which indicates the dynamic nature of the Metaverse. Changes to a physical object are mirrored in its virtual counterpart in the Metaverse because of their digital twins, which are linked to real-world objects. People can also change objects by interacting with them. Instead of just looking at digital objects in the Metaverse, users will be able to experience a place and interact with them [45]. The creation of new objects will require complex inputs and will demand high-quality user interaction with the objects in the Metaverse. The Metaverse poses three crucial challenges for effective user interaction, as depicted in Fig. 6:

**Interacting with existing objects:** Users' physical interactions with these virtual worlds are an important consideration [46]. For the Metaverse to persist, this is a fundamental challenge that must be overcome. When the user is unable to control the interaction, they will stop using it immediately. When a user is completely immersed in a virtual world and finds themselves unable to perform a task that they could do in the real world, they become frustrated and annoyed.

**Modifying existing objects:** As technology gets better and the real world keeps changing, the Metaverse objects will need to be changed to make them seem more real [47]. Realistic objects need more precise modelling algorithms, just like real faces. Even in the Metaverse, where scenes and avatars are always changing and interacting, objects have to be changed all the time to reach this level of realism.

**Creation of new virtual objects:** The Metaverse is a virtual 3D universe comprised of virtual 3D objects [48]. The Metaverse requires the creation of immersive experiences based on real-world artefacts to accomplish its objective of combining the digital and physical worlds. In the Metaverse, a lot of digital objects will need constant sensor inputs from their physical counterparts to produce this realistic immersive experience for the users. The Metaverse will also enable its users to create virtual objects by providing them with various tools and applications. As a result, it creates a huge requirement for bandwidth, which is a challenge to achieve with the present technology. From the above discussion, it is obvious that efficient user interaction plays a significant role in the creation, interaction, and modification of digital objects in the Metaverse. This requires massive input from real-world objects. 6G's URLLC and real-time processing abilities will aid in the building of a highly immersive 3D environment in the Metaverse.

### F. 6G FOR LOW LATENCY COMMUNICATION IN THE METAVERSE

Low latency communication is the capability of the communication network to deliver large quantities of data with minimal delay and high accuracy. These networks are designed to support operations that require access to rapidly changing data in real-time. Advanced technologies like self-driving cars, holographic telepresence, remote surgery, deep-sea and space tourism, and other AR and VR innovations are becoming part of the Metaverse [49]. For instance, we had been accustomed to virtual communication using Zoom, Skype, Microsoft Teams, and other platforms. Future developments in VR and AR are well on their way to making an office where people can talk to each other in a fully immersive way. This integration of advanced technologies into the Metaverse creates a huge demand for next-generation networks with enhanced bandwidth and latency.

The present network infrastructure cannot provide the bandwidth and latency required for the Metaverse and its applications. The capacity of current 5G networks to handle the IoE, holographic telepresence, collaborative robotics, and deep-sea and space tourism is limited. These applications require multiple terabytes of bandwidth as they depend on real-time inputs from the real world. From the discussion, it is clear that the Metaverse necessitates the highest network positioning accuracy and multiple terabytes of bandwidth. The 6G network, with its advancements like greater use of the distributed radio access network (RAN) and the terahertz (THz) spectrum to increase capacity and improve spectrum sharing, will provide effective and low-latency communication required for the Metaverse [50].

### G. 6G FOR COMPUTER VISION IN THE METAVERSE

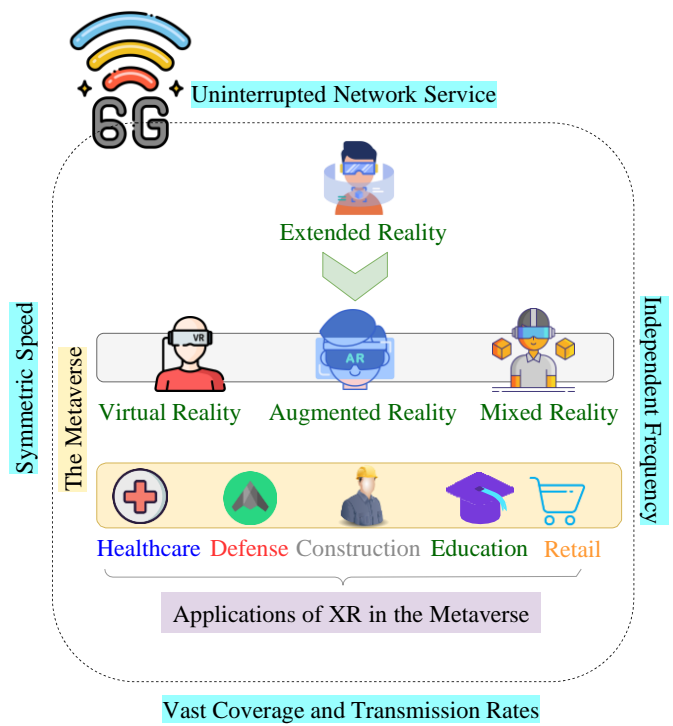

**FIGURE 7.** 6G for Computer Vision in the Metaverse

Computer vision is the study of how computers perceive and interpret digital images and videos. Computer vision encompasses all activities done by biological vision systems, including seeing or sensing a visual signal, interpreting what is being seen, and extracting complicated information in a form usable by other processes [51]. Using sensors, comput-ers, and machine learning algorithms, this multidisciplinary field replicates and automates essential parts of human vision systems. The objective behind computer vision is to develop artificial intelligence systems that can see and understand their surroundings.

In the Metaverse, computer vision plays an important role in enabling humans to experience the virtual environment, as depicted in Fig. 7. Through the use of digital avatars, VR and computer vision provide a near-to-lifelike experience in the Metaverse [52]. In order to connect to this virtual world, the user needs to use XR devices, which are built on the foundation of computer vision. XR applications rely heavily on computer vision. Visual information in the form of digital images or videos is often processed, analyzed, and interpreted with the help of computer vision and visual information. This helps people make effective decisions in the Metaverse. As a result of computer vision, VR and AR environments can be built that are more accurate, trustworthy, and user-friendly than their real-world counterparts. Human position tracking is a computer vision challenge that tries to figure out where people are located in an environment that is constantly changing. In the Metaverse, the healthcare, military, construction, manufacturing, education, and retail sectors will rely largely on computer vision. For example, doctors can improve surgical processes and study data from 3D scans in real time using computer vision. The computer vision will assist doctors in detecting, diagnosing, and treating potential diseases and enable them to examine patients from anywhere in the world [53]. Computer vision in the Metaverse will evolve at an accelerated rate, and even 5G cannot compete with the rapidly evolving technological requirements of the Metaverse's computer vision capabilities. The computer vision requires the continual collaboration of heterogeneous devices to provide immersive experiences for the users, which requires uninterrupted network service, and should provide symmetric uploading and downloading speeds for users to quickly upload all their content while concurrently downloading the content of others. 6G supports a higher number of device connections, which is very crucial for computer vision in the Metaverse for delivering its fully immersive services to customers [54]. The independent frequency, higher data transmission rates, and large coverage of 6G will enhance the QoS of computer vision in the Metaverse.

### H. 6G FOR HIGH TRANSACTION INTEGRATION/ SCALABILITY

To date, the Metaverse implementations used a centralized cloud-based approach for avatar physics emulation and graphical rendering. The centralized design is unfavourable as it suffers from several drawbacks caused by the long latency required for cloud access. Further deployments of Metaverses will also bring scalability issues to the physical layer due to the increased number of computing tasks mainly generated by extremely demanding applications. The traditionally deployed centralized architectures are unlikely to support a large number of Metaverses and their users, so the introduction of de-centralized Metaverse systems including frameworks and protocols is inevitable. There are several approaches that can be taken, starting with leveraging Mobile Edge Computing (MEC) technology. For example, [55] proposed the blockchain-based MEC architecture, where base stations allocate their computation and communication resources for providing video streaming and the use of a series of smart contracts enables a self-organized video transcoding and delivery service without a centralized controller. Using the MEC more efficiently will not fulfil the requirements in full, so the decentralized architecture will have to further distribute the communication and computational cost among different nodes present in the virtual space. The concept of de-centralizing the Metaverse applications was presented by authors of Solipsis [56] - a system that allows the adaptive streaming of 3D models including avatars, sites and objects in a completely decentralized fashion. In general, to overcome challenges related to a high number of transactions, massive resource demands and scalability concerns a novel framework should be proposed to address those emerging challenges for the development of future Metaverses. In such framework, the Metaverse Service Provider (MSP), which is a service provider that offers applications or services such as games, conferences or concerts should be able to get paid for provided services and in addition to this, the MSP should be allowed to negotiate with the Metaverse User (MU) to use MUs computational resources in return for discounts or rewards. The blockchain, which is provided by the MSP can contain all interactions between the MSP and MU in terms of transactions. The MetaChain [57] describes a similar concept that could serve basis for future deployments. In this particular proposal, the blockchain shards are used to allow the MUs to contribute their computational resources to the Metaverse application provided by the MSP. This is done in exchange for digital assets such as tokens or service access. With this approach, more users can be attracted to a particular Metaverse. However, attracting users to contribute resources is going to be particularly challenging. The reason is that the service provider will not be able to directly allocate the user resources to the shards. Sharding is so far one of the most practical solutions to achieve a scale-out system where the processing, storage, and computing can be conducted in parallel. As such, the capacity and throughput being linearly proportional to the number of participating nodes or the number of shards become possible, while preserving decentralization and security. The consideration has to be taken when creating shard-based systems as users (by human nature) will aim to maximize their profits and concentrate resources on the shards that pay more. Nevertheless, whichever form such framework will take, a pay-per-share protocol is required to off-load computational workload onto the Metaverse user devices.

### I. 6G FOR SECURITY AND PRIVACY PROTECTION, ELIMINATED CRIMINAL/HACKER ACTIVITIES, TRUST AND ACCOUNTABILITY

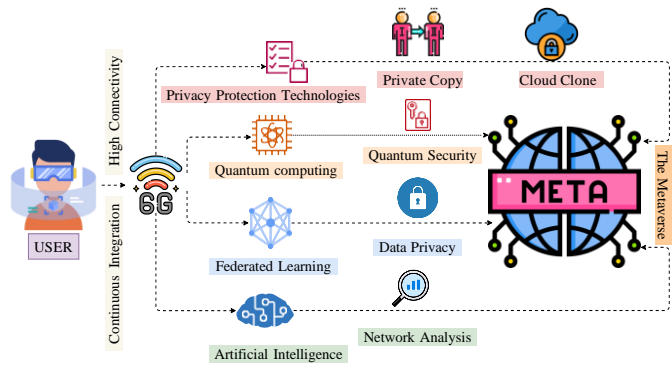

**FIGURE 8.** 6G for Security and Privacy Protection

Metaverses should offer their users an extraordinary immersive experience in virtual environments, such as entertainment games and smart cities, using its enabling technologies. The metaverse can track users' physical actions and physiological responses and may expose confidential information about their habits and physiological nature to third parties. If hackers get their hands on such sensitive information, it could lead to harassment and the theft of digital assets, which could make users lose faith in the security and privacy of the metaverse. These issues can be addressed by utilizing privacy-protection technologies like "Private Copy" and "Clone Cloud", as depicted in Fig. 8. The creation of private copies and clone clouds is dependent on high connectivity and continuous integration with the metaverse environment. The edge intelligence facilitated by 6G can support the needs of these technologies in the metaverse. The use of a blockchain-based digital twin wireless network and an edge computing federated learning architecture can further enhance the users' privacy and data security [8]. Together with 6G, AI can optimize connectivity while also enabling traffic prediction and improving security in the metaverse. To avoid information breaches, physical layer communication may use a machine learning-based antenna design. Machine learning and quantum encryption can also be used to protect the security of communication devices in the metaverse. The metaverse's security may be increased by using early warning systems and AI-enabled 6G to identify network anomalies. The use of distributed and federated AI in a 6G network also eliminates the necessity for data sharing across the metaverse devices, which preserves the privacy of the users.

### IV. ROLE OF 6G TECHNOLOGIES FOR THE METAVERSE

6G will play a key role in the Metaverse operation since such an environment requires pervasive connectivity for full-fledged and omnipresent Metaverse immersion. Essentially, very-high bitrates and ultra-low delay are crucial for a satisfactory Metaverse experience. An important factor on this performance is the smart management of connectivity resources/services, scalable infrastructure and very low latency communications. Therefore, Edge AI and cloud infrastructure are necessary for efficient and performant handling of relevant use cases in the Metaverse. Edge AI is an important enabler since it facilitates AI-driven optimized network management and minimizes delay with distributed and close-to-the-user computing paradigms. This technology will be compounded with the AI native design of 6G which will be embedded for numerous functions ranging from physical layer control to service management. Furthermore, the required flexibility and scalability for network and service environment requires moving towards cloud-native technologies which can also form telco clouds for more efficient and scalable Metaverse infrastructure in the backend.

In the cyber-physical domain, another aspect of the Metaverse regarding 6G will IoE and robotics play a key role. Additionally, 6G will have the essential toolbox to enable AR/VR, which is critical since the Metaverse will be the main vessel for AR/VR experience. An appropriate immersive experience in the Metaverse will be possible with those technologies enabled by 6G communication and computation functions. As a transversal technology similar to AI, blockchain can also help the distributed and open nature of the Metaverse and enable the transferability of digital assets which will be an important capability for the Metaverse use cases. A depiction of these technologies and their roles is provided in Fig. 9 and the summary of all related works is presented in Table 4.

### A. AI
#### 1) Introduction
Based on the combination of many advanced technologies, the Metaverse should be built to convey a groundbreaking immersive experience to users, in which AI has played a vital role in the foundation and development of the Metaverse regarding numerous aspects, including core services and applications. Besides the responsibility of ensuring the reliability of the Metaverse's infrastructure, AI can help developers in designing and building a smarter and more beautiful virtual world and further allows users to acquire hyperreal creation using built-in tools. In 6G systems, numerous challenging tasks and applications can be solved and enabled by advanced ML algorithms with different learning mechanisms (i.e., supervised learning, unsupervised learning, and reinforcement learning) to achieve high performance and low latency. Especially, DL with the advantage of effectively learning complex patterns from practical large and messy datasets will be the key technology to polish many aspects of the Metaverse, from the intelligence of AI agents and virtual assistants (a.k.a., chatbots) to the visual quality of 3D worlds [58]. Indeed, the presence of AI in the Metaverse can be realized in the interactions between a user (represented by an avatar) and other objects (e.g., non-player characters) by automatically analyzing sensory data for multiple tasks, such as speech recognition and understanding, facial expression analysis, body movement tracking, and gesture recognition. Besides,

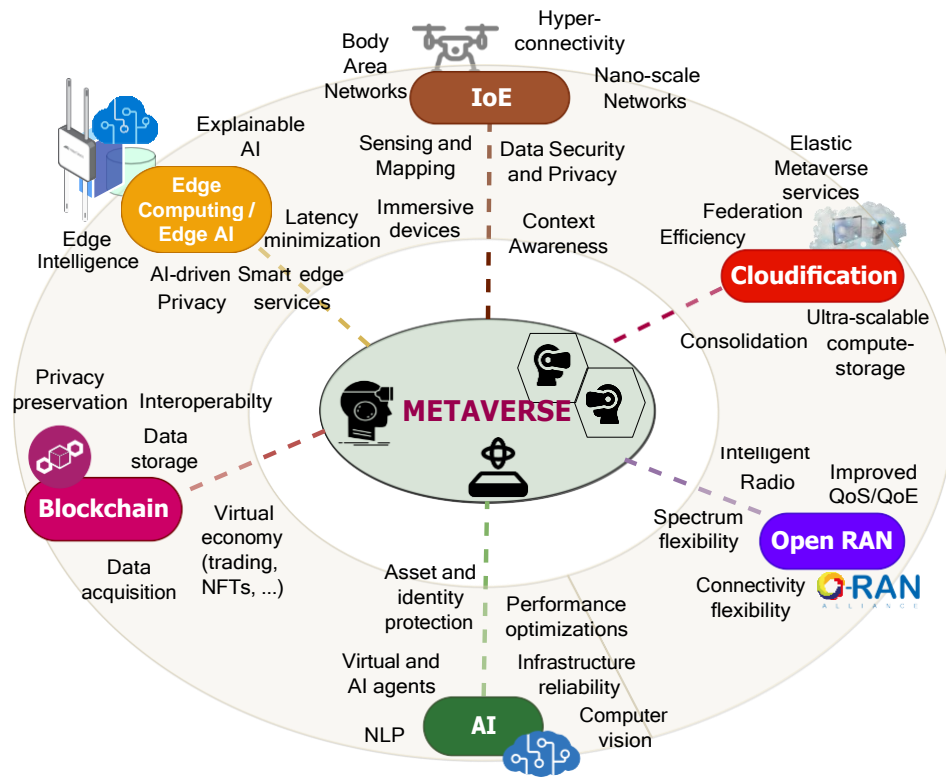

**FIGURE 9.** 6G key technologies and their roles for the Metaverse.

AI can be applied to preserve users' identity and their digital assets from cyberattacks, especially in the scenario in which the Metaverse is built with a cross-chain bridge.

2) *How 6G AI can help, on which features*

In the Metaverse, natural language processing (NLP) plays an important role to deploy intelligent virtual assistants (including chatbot) [59], which helps the Metaverse comprehensively understand what users are typing and talking, from simple sentences to complicated and long conversations, over unlimited, to smooth user interaction accordingly. Empowered by AI with ML and DL algorithms, chatbots can immediately respond to the users and adapt to an environment with reinforcement learning to consolidate operation and improve the performance of an overall virtual assistant system [60]. In the NLP domain, language modelling aims to predict linguistic components in sentences and paragraphs by mining syntactic and semantic relations of words and phrases, which is commonly developed for machine translation and text recommendation. Several advanced language modelling methods have exploited DL with RNN, LSTM, and CNN architectures to improve the overall system efficiency and addressed many fundamental tasks [61], such as identifying long-term dependency in long sentences in complicated scenarios, recognizing hyphenated words, misspelt words, suffixes, and prefixes. Especially, language modelling should be taken into consideration with different popular languages, such as English, Chinese, Spanish, and French [62], [63] to attract as many as possible users from over the world to join the Metaverse. Some advanced structures in deep networks, such as bidirectional LSTM, bidirectional gated recurrent unit (GRU), and channel-wise attention connection, have been leveraged to deal with some challenging problems, such as sentiment analysis, question type classification, and answer identification with multiple sentences [64], [65], which accordingly improved readability, interpretation, and rationality of virtual assistant agents. Some other specific AI-based NLP tasks (e.g., context retrieval, semantic notation, and named entity recognition) can be considered to uplift text-based and speech-based user interactive experiences in the Metaverse.

Commercial headset devices with VR/XR technology have been designed to bring 3D viewing experiences to users, including high video quality (i.e., high resolution and high frame rate) and wonderful wearing comfort thanks to the advancement of AI. In [66], an eye fixation prediction method was introduced for gaze-based applications (e.g., video rendering and content visualization), in which a DL framework with hierarchical CNN architectures was exploited to process different data types, including VR images, gaze data and head data collected by wearable sensors. Some recent works have studied advanced ML and DL algorithms to precisely identify periodic behaviours of VR gear's user (e.g., gaming controllers and head-mounted display) for automatic identity authentication and health issues detection [67]. Some deep networks with CNN architectures have been designed to assess the quality of images/videos (e.g., color saturation, brightness, resolution, and frame rate) displayed on the screen of VR devices, and then automatically adjust screen settings to optimize the visualization based on video contents and user conditions [68].

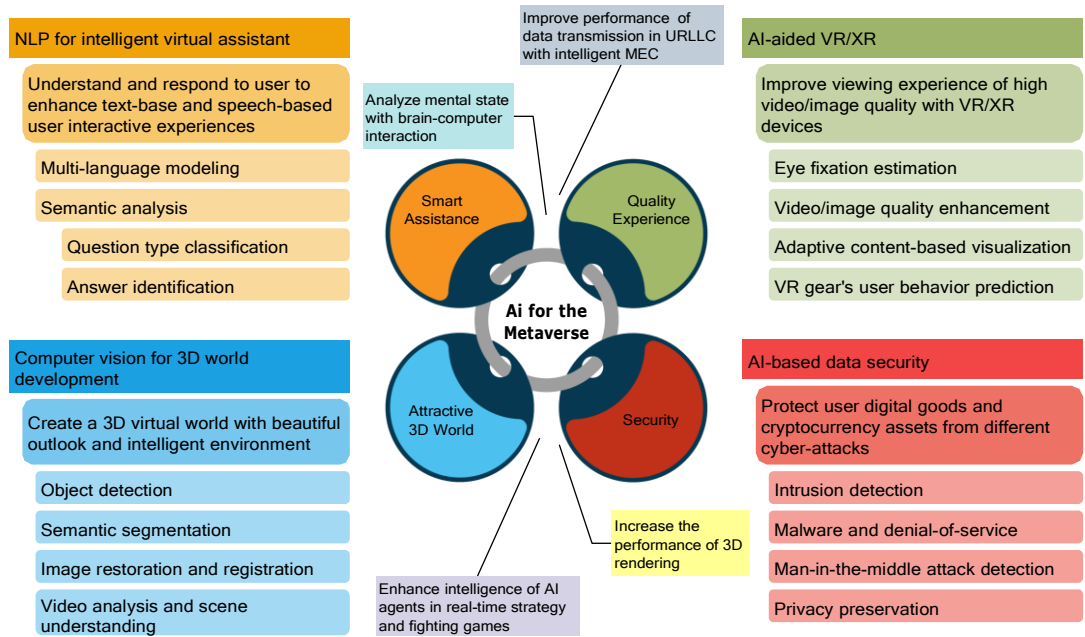

**FIGURE 10.** The roles of AI for the development and advancement of the Metaverse.

Along with the VR/XR technologies, computer vision is one of the most important sectors to build a beautiful virtual world in the Metaverse and enable it to be more intelligent from the user's viewpoint with the adoption of AI, especially DL in a few years [69]–[71]. Many sophisticated CNN architectures have been designed for different fundamental image processing and computer vision tasks, such as object detection, semantic segmentation, and scene understanding [72], [73]. For instance, atrous convolution was introduced by DeepLab [74] for semantic segmentation to capture more meaningful features by enlarging the receptive field of kernels to enhance the learning efficiency of a deep network while obtaining a small network size. Static and dynamic objects can be detected and located in the virtual worlds accurately by several recently advanced DL models to provide useful information to users and be synthetized for higher-level tasks like scene understanding and detailed captioning. Some image/video quality distortion problems, such as blurring, noise, and frame corruption, can be addressed effectively by AI technology to guarantee the high-class visual perception of a user when experiencing the Metaverse [75]. In addition, the activities, including single actions and interactions, of users and non-player characters in the Metaverse can be automatically detected and recognized by AI-powered human pose estimation and action recognition methods [76]. Some convolutional networks have exploited cutting-edge layer structures, such as dense connection, skip connection, and attention connection, to estimate complex human poses and classify grouped activities while handling other challenges like varying viewpoints, object occlusion, and complicated actions in practice. For example, generative statistic models and hybrid LSTM-CNN architectures are suggested in [77] to precisely examine pose transition in the spatiotemporal domain, thus increasing the accuracy of action recognition and activity classification.

To preserve the Metaverse from different cyberattacks, especially protect users' digital goods and cryptocurrency assets, many advanced ML algorithms and DL models can be deployed in multiple layers (e.g., network and services layers) of the Metaverse's platform for intrusion detection [78]–[80], in which various malicious attacks can be automatically and accurately detected and classified to immediately provide an efficient security solution. In [81], a holistic security method with sufficient assurability and explainability was proposed to quickly and sequentially detect abnormalities and time-dependent abnormal events in IoT systems and software-defined networks, in which zero-bias neural networks are transformed into performance-assured binary abnormality detectors to increase detection accuracy while presenting the lowest latency based on false alarm constraints. In the effort to exploit DL denoising autoencoder (DAE) for many fusion security problems, many variants of DAE, including stacked autoencoder, stacked sparse autoencoder, stacked noise autoencoder, stacked contractive autoencoder, deep belief network, were benchmarked in for performance comparison with different practical intrusion detection datasets [82]. Reinforcement learning (RL) with the capability of learning environmental factors to adapt learnable parameters was also exploited to deal with different types of cyberattacks (such as malware, denial-of-service attack, and man-in-the-middle attack) [83]. In addition, privacy preservation in the Metaverse should be uplifted comprehensively with the help of AI to ensure that there are no leakable risks and threats to users' big data in the virtual world. For instance, a privacy-aware and asynchronous DL-based method was introduced in [84] to maintain the confidentiality of data among different collaborative data collection sites. In [85], an optimal centralized privacy-preserving aggregate mobility data release mechanism was proposed to minimize the data and information leakage, in which deep RL models and the Asynchronous Advantage Actor-Critic algorithms are combined to optimize the privacy-preserving method. The above-mentioned privacy-preserving DL-based methods can be recommended for the Metaverse to combat information leakage threats and adversary attack effectively.

3) Summary

In the Metaverse, AI has presented a plentiful foundation and development in numerous aspects and helped to construct a more beautiful virtual world with intelligent and secured services, thus bringing a wonderful experience to users. Several advanced ML algorithms and DL architectures have been deployed to take care of the comfortableness of VR users, and the interaction between users with virtual assistants, and automatically provide useful information about the virtual worlds to users. Besides some popular domains like NLP and computer vision, AI has great potential for deployment in other sectors: protecting users' digital assets from hackers, early detecting intrusions for data security and privacy preservation, improving the performance of URLLC with intelligent MEC, enhancing the intelligence of AI agents in real-time strategy and fighting games, and analyzing mental state with the brain-computer interface as illustrated in Fig. 10. Although some advanced ML and DL models can conduct a high performance in many detection and classification tasks, they represent black boxes that lack the capability of explainability and interpretability. Therefore, there remains room for AI research and development in the Metaverse.

### B. BLOCKCHAIN
1) Introduction

In the Metaverse, data privacy and virtual asset (e.g., cryptocurrency and NFT) security of users should be guaranteed as the top priority. In this context, blockchain technology represents a promising solution with many unique features at once, for example, decentralization, transparency, and immutability. Fundamentally, blockchain is an innovative technology that permanently records transactions in a decentralized and public database so-called a ledger [86]. Although all transactions are transparent (i.e., being available to check by anyone), the decentralized recording system of blockchain is very difficult to fool or control. Some blockchains like Ethereum and Solana are programmable through smart contracts with different consensus mechanisms, such as proof-of-work and proof-of-stake, which can meet high-security requirements of e-commerce platforms and enable the revolution of the digital ecosystem in the Metaverse, especially supporting virtual asset payment and trading activities. A smart contract on the blockchain could be used to establish the ownership of any digital object, such as artwork and music, over NFT specialized by unique and nonreplaceable (i.e., no one else can claim the ownership of that digital product on the blockchain even if they have a copy version on computers). The role of blockchain in the Metaverse relies on ensuring data privacy and security, enabling seamless and secured data sharing, data interoperability and integrity with some common applications and services, such as decentralized finance and NFT market [87]. Besides that, blockchain allows digital goods to be tradable safely in a virtual world and enables the connection of physical objects to the Metaverse over NFTs. Notably, if two virtual worlds are interoperable, the blockchain has to authenticate the proof of ownership of digital goods in both virtual worlds. Indeed, blockchain bridges the real world and the virtual world besides playing as the gateway for users to access the Metaverse.

2) How 6G BC can help, on which features

Data acquisition is one of the most fundamental processes to build the virtual world in the Metaverse, which collects big data from different modalities. Notably, the sensitive data collected from users to train AI models for several special modules (such as decision-making of virtual assistant, recommendation system, digital product development, and automated market maker) in the Metaverse should be secure. For secure and large-scale environment data acquisition, the work in [88] proposed a blockchain-based system which is specialized by one valuation layer to assess the quality of acquired data, one consensus layer to encourage and incentivize high-quality data acquisition, and one ledger layer to record transactions and qualified environmental data. In [89], a blockchain-based efficient data collection and secure data sharing mechanism was introduced for reliable industrial IoT systems. This mechanism has exploited the Ethereum blockchain to maximize the amount of acquired data and the deep reinforcement learning algorithm to obtain highly secure and reliable shared data. To guarantee the users' privacy in crowdsourcing systems, Li *et al.* [90] designed a blockchain-based decentralized framework for data collection and sharing. There were three standard smart contracts on blockchain executed for the whole process of data acquisition to achieve such crowdsourcing information as task posting, receiving, and assignment. The proposed method was implemented and verified on an Ethereum test network with real-world data, which demonstrated usability, feasibility, and scalability to be suitable for distributed crowdsourcing systems. Although blockchain technology can ensure highly secure and reliable data supplied to the Metaverse, its drawback is low latency due to the complicated and distributed nature of processing transactions with smart contracts and consensus mechanisms like PoW. Besides, the high transaction fee is also a realistic barrier for a low-income user to experience the Metaverse.

In a large-scale Metaverse platform, data storage should be taken into consideration seriously because of the high velocity, big volume, and complicated variety of big data from a plentiful number of applications and services deployed in virtual worldds [91]. There exist many underlying risks, such as leakage, tampering, and loss if the Metaverse is built on a platform with centralized storage systems. Some sensitive data like biometric login data of the user (e.g., face and touch identification on iPhone) can become the target of cyberattacks to steal virtual assets. To overcome the above-mentioned issues of centralized systems, the work in [92] proposed a large-scale secured IoT data storage scheme by exploiting blockchain miners to manipulate IoT data stored in distributed hash tables (DHTs). In the blockchain system, a certificateless cryptography scheme was applied to reduce redundancy in traditional public key infrastructure and authenticate IoT devices, where the generated public key pairs are broadcasted to all devices with verification done by the blockchain miners. In [93], the time-series data in the Metaverse was stored in a locality-aware auditable decentralized storage ecosystem that was designed and managed thanks to the advancement of blockchain technology. Some data storage systems with recovery functionality have been developed to effectively address multiple problems, such as low integrity, high cost, and easy tempering. Liang et al. [94] introduced a secure blockchain-based data storage scheme, wherein the incoming data packets are verified with smart contract and consensus mechanism, and then checked to early detect any threats before being stored on a decentralized system. Notably, when distortion occurs to the stored data, multiple nodes in the blockchain network can repair it successfully.

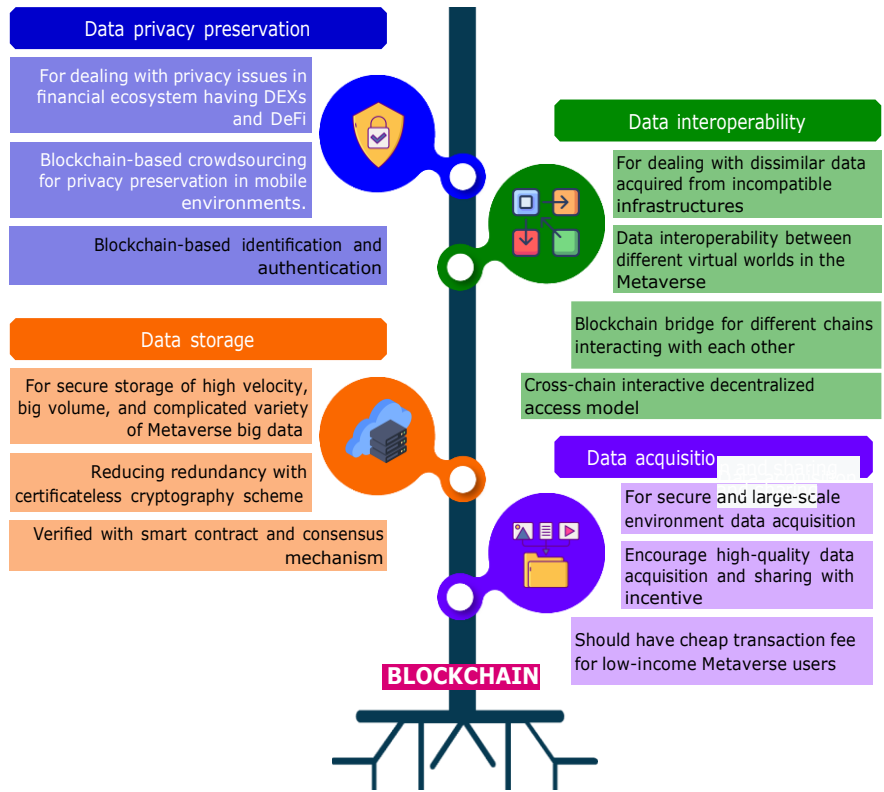

**FIGURE 11.** The roles of blockchain for ensuring the security and privacy of data acquisition, data sharing, data storage, and data interoperability in the Metaverse.

As the mixture of numerous digital realms, the Metaverse demands manipulating and processing the big data that is acquired from incompatible infrastructures for different purposes, in which the standardizations of data for different applications and services in the virtual worlds are dissimilar. This reveals a serious concern about data interoperability when expanding the Metaverse with an interconnection capability among different virtual worlds. To ensure the interoperability between different virtual worlds in the Metaverse, building a cross-chain protocol or an inter-blockchain bridge becomes a promising solution in many specific domains like healthcare and e-commerce [95]–[97]. A blockchain bridge is a protocol connecting two economically and technologically separate blockchains (such as Bitcoin, Ethereum, Avalanche, Solana and Polygon) for interactions and acts like a physical bridge linking the ecosystems of one blockchain with another. As a result, blockchain bridges enable what is called interoperability means that digital assets and data hosted in Metaverses built on different chains can interact with each other [98]. Besides, blockchain bridges allow users to access new protocols on other chains and encourage collaboration between developers from different blockchains, thus promoting a virtual economy in the Metaverse. A novel blockchain framework, namely BiiMED, was introduced in [95]to uplift the data interoperability and integrity in electronic health records (EHR) sharing systems. The proposed framework facilitated the medical data on EHR systems between different medical providers and healthcare institutions with a decentralized trusted third-party audior for interoperation validation. Some recent cross-chain protocols [99], [100] have been introduced to interconnect multiple blockchains for secure data utilization and management while obtaining full interoperability. In [99], a cross-chain interactive decentralized access model was designed with a gateway to reading the information on multiple chains and route cross-chain transactions, and a notary network with an interplanetary file system and BigchainDB to verify and confirm each transaction based on a voting mechanism. Such kinds of cross-chain protocols allow users to easily buy, sell, and trade virtual assets among different digital worlds without any intermediate tools, and consequently encourage the adoption of the Metaverse. Along with interoperability, data integrity has also received much attention in the Metaverse, in which blockchain technology was considered to verify and protect data integrity in decentralized cloud computing systems [101], [102].

In the Metaverse, a user can freely interact and trade virtual goods (including cryptocurrency and other virtual assets like NFT) with a virtual assistant and other users via decentralized exchanges (DEXs) integrated into the Metaverse to promote the development of decentralized finance (DeFi). As an ecosystem of financial applications built on blockchain networks, DeFi enables easy access to financial services, facilitates traditional financial systems, and has a modular framework with interoperability with public blockchains. Recently, GameFi, a fusion of words game and finance, refers to play-to-earn blockchain-based games with economic incentives to players, which is being developed and integrated in the Metaverse. A GameFi ecosystem uses cryptocurrency, NFTs, and blockchain technology to create a virtual gaming environment, where various GameFi ecosystems built on different chains can be involved in the Metaverse owning to chain bridges. In this context, it arises many privacy issues (e.g., the leakage of user identity and other personal information that can be stolen for illegal purposes) can be effectively handled by blockchain technology with immutability [103]. In a blockchain-powered Metaverse, third-party intermediaries are not permitted to manipulate the data of other parties. In [104], a blockchain-enabled crowdsourcing approach was proposed to deal with privacy preservation in mobile environments, where users can access the Metaverse using mobile devices. In secure 5G and 6G communication networks [105], blockchain was exploited to minimize privacy breaches by completely integrating authentication mechanisms with blockchain-based identification systems.

### 3) Summary
With the distinctive features of decentralization, immutability, and transparency, blockchain technology has promoted the development and advancement of the Metaverse, where it has played an important role in any Metaverse platforms with some great contributions in terms of many technical aspects, including data acquisition, data storage, data interoperability, and privacy preservation. Besides ensuring the privacy of sensitive information and security in trading activities (e.g., buy/sell cryptocurrency, NFTs, and other virtual assets), blockchain has shown great achievement to revolutionize user's immersive experience, boosting the economic growth, and attracting new users to the Metaverse via numerous blockchain-aided applications and services supplied in the virtual worlds. However, it remains several challenging issues to concurrently attain security, scalability, and decentralization when the Metaverse must serve a huge number of users and a rapidly increasing number of transactions to process. Consequently, many research topics to optimize blockchain for the Metaverse should be continuously exploited in the future, such as consensus algorithms, blockchain interoperability, smart contract, and network management.

### C. EDGE COMPUTING AND EDGE AI
#### 1) Introduction
The Metaverse is envisaged to map and simulate all our daily life activities in cyberspace at a huge scale while enriching such mapping with an immersive and interactive user experience. Cyber-physical and digital twin applications will also be integrated with the Metaverse application to offer realistic cyber representations of the physical world. In the ICT infrastructure, there will be the Metaverse engine which performs computations required to run virtual universe simulations carrying out computationally heavy tasks such as collision detection in the virtual universe and computation of 3D physics, and also other aspects of virtual universe that demand high computational power [106]. The Metaverse is striving to connect billions of users and create a shared world where virtual and reality merge [8]. Therefore, users interact in the physical-virtual world with the characteristics of diversification of information, identities, and modes under the requirements of ultra-low latency, massive resource demands, interoperability between applications, and security and privacy issues [107].

The promised Metaverse operation will require extremely low latency with highly elastic and omnipresent compute and storage resources. The latency and processing challenge for the Metaverse is in line with what is expected with 6G edge computing realization: For the Metaverse extended-reality computations to be offloaded, the entire process must be shortened so that input from the user device, a network trip, processing by the service, a return network trip and drawing the output on the user device fits in the 20ms time taken by a single network trip today [108]. Cloud-based processing for Metaverse operation can be unfavourable as it suffers from several drawbacks caused by the long latency required for cloud access, such as low-quality visualization in XR. 6G enables real-time, ubiquitous, and ultra-reliable communications for massive Metaverse devices with support for device mobility, which can reach 1020 Gbps [109]. To this end, Fog

Computing [110] and Mobile Edge Computing [111] have been proven effective to tackle the issues faced by cloud-based systems, by moving the computational heavy load near the end-user and distribute it among edge devices; such approach can significantly reduce the latency and optimize the system performance. Furthermore, there is the cost-benefit: such an approach it would drive down the cost of XR devices and allow mass adoption. Verizon has estimated that any more than 20ms of motion-to-photon (total stack) latency causes many users to become nauseated; for comparison, well-built wireline broadband networks today typically have 20ms of network latency alone, and typical LTE latencies are 3x higher. Therefore, edge computing is an important technology. For instance, Zhang et al. [112] introduced the MEC into the Metaverse to improve the quality of users' experience. Xu et al. [7] discussed the potentials of AI, Edge computing and blockchain for ubiquitous, seamless access to the Metaverse. Similarly, Lim et al. [113] present the infrastructural architecture required for the Metaverse with a special focus on the convergence of edge intelligence and the infrastructure layer of the Metaverse.

6G-enabled edge intelligence opens up a new era of Internet of Everything and makes it possible to interconnect people-devices-cloud anytime, anywhere. In this context, industry, and academia have developed a new learning paradigm, Edge Artificial Intelligence (Edge AI) [114], which allows AI models to be deployed on devices and perform real-time data processing and model inference. 6G mobile communication technology provides edge AI with lower latency, more stable network connection, and more secure network architecture. Edge AI with 6G is expected to be applied to solve problems such as high bandwidth and high connection density in the Metaverse. However, the Metaverse still faces many challenges, such as users' privacy, network latency, and resource allocation issues. Moreover, the Metaverse places higher demands on the current edge AI architecture. As mentioned above, 6G edge intelligence has the advantages of low latency, computing offload, and high performance [115]. Overall, the application of 6G-oriented edge intelligence has the benefits of balanced data storage, efficient data transmission and high reliability.

2) How 6G EC and Edge AI can help the Metaverse

As noted above, a high-speed and low-latency network connection and ubiquitous access to services is an important foundations for improving the user experience in Metaverse. Otherwise, issues such as visual jitter or delay and other undesirable phenomena might lead to the subpar performance of Metaverse. In that regard, to reduce network latency, an incentive mechanism framework for VR services was proposed in [116], which uses perceived quality as a criterion for measuring immersive experience and effectively evaluates the immersive experience in the Metaverse. [117] presents a novel MEC-based mobile VR delivery framework that is able to cache parts of the field of views (FOVs) in advance and compute certain post-processing procedures on demand at the mobile VR device. Jiang et al. [118] found that coded distributed computing (CDC) can improve the latency problem in the Metaverse and proposed a CDC and dual blockchain distributed collaborative computing framework.

However, the computing, communication, and storage shortage will seriously affect the user's immersive experience. For the resource allocation problem, a new blockchain-based framework called Metachain was proposed in [87] which uses Stackelberg game theory analysis to propose an incentive mechanism, i.e., users obtain corresponding rewards by providing resources to blockchain shards. Based on the intelligent 6G edge network, a machine learning framework was proposed in [119] for decentralized learning and coordination of edge nodes to improve resource allocation strategies.

For Edge AI and its applications in 6G, there are various challenges which are investigated by the research community. Edge AI paradigm and its applications still have the following issues that need to be optimized [8]:

– High Latency: Since edge AI generally involves thousands of remote devices and needs to transmit and process massive amounts of data [120], [121], the high latency issue in the current network environment has always been one of the bottlenecks hindering the wide application of edge AI [122], [123].

– Fragile Stability: In edge AI, the training of large-scale models often requires powerful computing power and stable network connections, especially the training of large language models [124]. However, the current network environment is only suitable for the training of small-scale models [125]. This is due to the fragility of the network connection leads to the failure of large-scale model training.

– Low Security: The current network architecture no longer meets the security needs of thousands of remote devices connecting to cloud servers today [120]. Furthermore, the openness of the network further challenges the security of the current network architecture.

These issues are expected to be exacerbated with the utilization of Edge AI in 6G for the Metaverse applications. For instance, in [8], Chang et al. propose a self-balancing federated learning-based Metaverse framework to address the statistical heterogeneity faced by edge-cloud architectures. Besides, in [126], Lu et al. proposed a blockchain-based digital twin wireless network (DTWN) edge computing federated learning framework to solve the problem of user privacy data security.

3) Summary

The integration of edge computing and realization of edge AI in 6G will provide various capabilities as well as challenges for the Metaverse. The key benefit is related to latency minimization needed for superb Metaverse user experience and pervasive services. Similarly, the inherent Edge AI support in 6G will also serve the Metaverse for smart edge services leading to better Metaverse services and device simplicity and flexibility. However, the potential benefits of 6G edge technologies should be supported with relevant research for improving on the aspects such as smart resource allocation, security, and privacy-preserving AI techniques.

### D. 6G OPEN RAN

1) Introduction

Radio Access Network (RAN) is a very important component of a mobile communication system that can link individual devices like mobile phones or terminals to other parts of the network using cellular radio. RAN coordinates the resource management in the cellular network across the radio sites. RAN can send the signal from a mobile device that is connected wirelessly to the core/backbone network to several endpoints in the wireless network, thereby, enabling the signal to travel along with the traffic generated from other networks. A RAN will typically comprise of base stations, that can transmit and receive signals to and from mobile devices in the network. The signals are then digitized in the RAN-based station and are connected to the network. RAN contains radio units (RU), distributed units (DU), a centralised unit (CU), and the RAN Intelligent Controller (RIC) for the creation of an effective mobile communication system. RAN is very important for meeting the low latency and high-speed internet connectivity requirements of real-time applications [127].

RAN requires manual intervention if any network issues arise in software or connecting devices. Pointing out the cause and the origin of these issues in the network by the mitigation experts is difficult as RAN is black-box in nature. The process involved in the mitigation of these network issues requires significant cost and time, subsequently affecting the overall quality of the network. This necessitates the creation of open, intelligent, virtualised, and fully automated interoperable RAN for the next generation 6G networks [128]. Open RAN (ORAN) is one such technology that integrates AI to optimize radio resources and also automates the management and operations of infrastructure. 6G ORAN integrated with AI can be used to implement Self-Organizing Networks (SON) and Radio Resource Management (RRM) solutions that improve network coverage, capacity, handover, and interference. They could also be used to increase the spectral efficiency of massive MIMO systems by optimising their performance. AI/ML can also enhance the user experience through VoLTE/video quality optimization, terminal anomalies detection, and other Quality of Service/Quality of Experience (QoS/QoE)-based use cases [129]. The use of ORAN gives mobile network operators (MNOs) the flexibility to provide 6G connectivity in a cost-effective, secure, and energy-efficient way. The openness of ORAN also enables the MNOs a unique ability where all the vendors can share the RAN functionalities [130]. As a result, it avoids vendor lock-in by replacing vendor-proprietary interfaces with a fully disaggregated RAN based on open standards.

2) How 6G Open RaN can help the Metaverse

The users navigate in the Metaverse frequently with the help of technologies such as AI, XR, digital twins, IoT, etc. As a consequence, the Metaverse demands continuous connectivity with sensors, hardware devices, and many other peripherals for providing high-quality and immersive services to the user. Any disruption in the network connectivity of these devices will cause the users extreme discomfort and make them feel that the surroundings are out of their control [131]. The standards for the Metaverse are substantially more demanding than those for the vast majority of internet applications in the present day. The current capacity of MNOs to handle the network requirements of devices connected to the Metaverse is rather questionable. This presents a challenge in the adaptation of the Metaverse. To solve these issues ORAN in 6G is a potential solution. ORAN in 6G with its AI, automation, and fully disaggregated open standards will enable the Metaverse to be cost-effective, secure, and energy-efficient way.

Let us consider the application of the Metaverse in the healthcare domain. The Metaverse allows healthcare professionals to have better interactions with patients who are in different demographic locations, such as viewing a three-dimensional model of the human body while discussing diagnoses and treatments. This would allow doctors to simulate the effect of a proposed treatment on a patient's body before its application, creating a more personal and informative experience than is currently possible with two-dimensional images displayed on a screen. VR, AR, and MR technologies are currently being used for medical training and surgical procedures, These enabling technologies of the Metaverse demand reliable connectivity. If any failure of software or hardware occurs in the network at the time of medical intervention it will lead to serious catastrophic situations. ORAN in 6G enables devices to relay on multiple MNO, so, this will ensure the medical devices connected to the Metaverse with much reliable connectivity. The remote medical surgeries supported by the Metaverse require real-time insights. The network supporting these devices must be faster in recovering from the related issue and failures. The ORAN in 6G will provide the Metaverse with zero-touch network and service management capabilities which will automatically resolve the raised issues related to the network faster than the traditional RAN. The vital monitoring devices connected to the Metaverse require a latency-free and cost-efficient network. These devices connected to the Metaverse will be greatly benefited by ORAN service management and orchestration platform in 6G. ORAN service management and orchestration platform in 6G is an intelligent automation platform that applies automation reduces the complexity of networks, improves network performance, and enhances the customer experience in the Metaverse which minimizes ORAN operational costs, as depicted in Fig. 12.

In the Metaverse, the possibilities of what can be created and purchased are nearly limitless. Users can purchase avatar skins, hairstyles, clothing, and accessories, as well as virtual land and property. Cryptocurrency and digital wallets will play a role in the Metaverse payments. Blockchain-based cryptocurrencies in the Metaverse or a crypto wallet are required to store and transport digital assets purchased in the Metaverse as well as between the virtual worlds. Digital wallets will be an alternative payment method that enables users to purchase digital goods securely. Thus the number of transactions occurring in the Metaverse will be limitless. Any breach or a critical update to the network will interrupt or halt these transactions and may affect the QoS/QoE of the customer in the Metaverse. ORAN in 6G will be less dependent on hardware which will reduce the risk associated with automated upgrades or isolated security breaches. The enhanced modularity available with open interfaces makes it easier for operators to serve the Metaverse towards a continuous integration/continuous delivery of services. Every trade or purchase that occurs in the Metaverse is recorded as a transaction, which results in huge network traffic because the data is to be stored in multiple peers. ORAN in 6G helps the Metaverse in better traffic management and also determines where to send traffic across the network. ORAN in 6G and AI enables the Metaverse to predict network conditions, such as congestion, so the controller can find an optimal path to send traffic over. This provides the users of the Metaverse with valuable insights about the network.

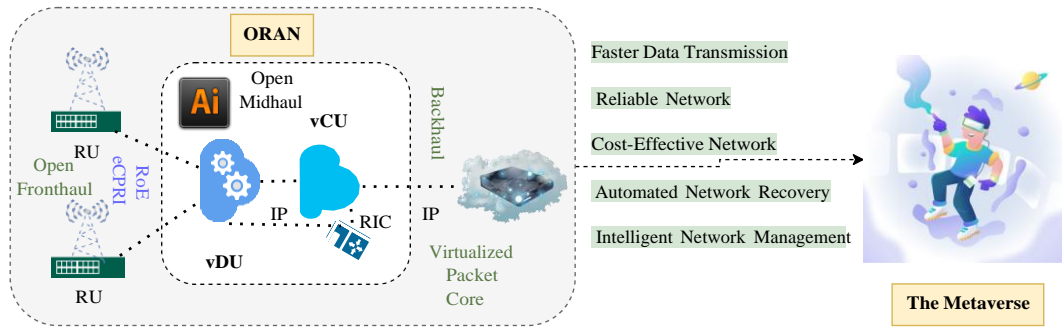

**FIGURE 12.** The role of 6G Open RAN for the development and advancement of the Metaverse.

3) Summary

ORAN in 6G with features like openness, better security, enhanced resource sharing, improved traffic management, and zero-touch network and services management provides the Metaverse with a network that is faster, reliable, cost-effective, automated, and intelligent. This is will help the Metaverse applications and services to be real-time. ORAN in 6G will help the users in the Metaverse with high-quality immersive experiences. The issues related to network software updates or threats will not affect the transactions in the Metaverse as ORAN in 6G is secured and depends less on hardware compared to the traditional RAN. ORAN in 6G allows AI to easily analyze the network and provide valuable insights for the Metaverse to persist. Though ORAN in 6G provides better network capabilities to the Metaverse it still faces challenges related to widespread adoption, technical support difficulties, system integration problems, and security risks.

E. 6G CLOUDIFICATION AND CLOUD NATIVE TECHNOLOGIES

1) Introduction

A key aspect of 6G networks will be the cloud-native design of the overall ecosystem. With the actual realization of the Metaverse, the cloud, infrastructure, and telecom companies will have to provide a fully immersive Metaverse experience challenging servers with a 100x harder compute task than an AAA game server hosting Fortnite today, and telecom access networks facing 24x more traffic in a decade [108]. To address these compute-storage requirements, the State-of-the-art Metaverse architectures rely on a cloud-based approach for core Metaverse functions such as avatar physics emulation and graphics rendering computation. Specifically, XR places extraordinary demands on networks with native cloud design of 6G networks, on-board computation capability to eliminate external computing device dependency can still be delivered on simpler, lighter, and cheaper end-user devices if computationally intensive tasks can be offloaded to a cloud computing instance.

The Metaverse leads to a clear need for cloud computing, considering the amount of storage and processing required to support a virtual reality universe: compute, storage and network loads [132]. As more performance and details will be demanded, remote cloud-based computers will become a necessary cost-effective way to solve that problem. The cloud computing technologies will be heavily exploited in two dimensions: First, by the Metaverse providers themselves built whether with private data centres or managed services. Due to their advantages, these compute- and graphics-intensive systems will be built on public cloud providers. Another option is to provide on-demand access compute and storage using pay-as-you-go models which can be done by public cloud providers with points of presence distributed globally. However, there is also the latency dimension: navigating the Metaverse smoothly through VR technology depends mainly on the network latency. As VR technologies are delay-sensitive and required very short latency, communicates with the Metaverse servers plays a pivotal role that leads to telco clouds where this concept is embedded in the telco network itself.

For example, the validation of Non-Fungible Token (NTF) trading transactions requires tremendous computational power. This challenge is also valid for other Metaverse applications such as data processing of digital twin applications or AI-enabled services like storytelling and recommendation services that empower the virtual universe simulation [106]. Current state-of-the-art Metaverse implementations perform the computational on the cloud, which may limit the simulation capacity, and increase access latency. Moreover, there are several independent and fragmented Metaverses that rely on different hardware and software technologies. Since the service providers would like to exploit the advantages of controlling users' Metaverse data in their servers, we may end up with isolated Metaverses rather than a universal Metaverse. Additionally, due to capacity limitations, the number of users that can access each region may be limited by the cloud service provider's local computational and communication capacity. Such limitations defeat the purpose of a virtual world, which is supposed to accommodate avatars as much as the real-world location can physically accommodate people.

Mobility support is also crucial since Metaverse will be a pervasive experience. Cloud can also help there as proposed by [106]. In this context, they propose a distributed architecture that can achieve a universal Metaverse, and solves the computational bottleneck. The advantage of layered architecture is twofold. Firstly, the users control their data, which enables organizations to access a universal Metaverse, rather than multiple separated Meta- verses. Secondly, the computational bottleneck is resolved by distributing the computational cost of heavy tasks.

2) How 6G cloudification can help the Metaverse

The real-time interactive nature and high demands on data storage, streaming rates, and the processing power of Metaverse applications will accelerate the merging of the cloud into the network, leading to highly distributed tightly-integrated compute- and data- intensive networks becoming universal compute platforms for next-generation digital experiences [133]. For instance, Google Stadia [134] and Nvidia GeForce Now [135] instead offload such rendering tasks to a remote compute cloud—allowing the highest level of quality on weaker devices such as smartphones. less latency- and loss-tolerant (to provide satisfying responsiveness to inputs). To an even greater extent than AAA video games, VR and MR are highly computationally intensive.

3) Summary

Cloud computing technologies and their adoption by telecom operators as telco clouds and cloud-native design in 6G have important implications for the Metaverse. First, they allow elastic Metaverse services which can be dynamically deployed and provisioned. Moreover, the Metaverse is expected to be a federated entity where different service providers, applications and users are present. Cloud computing enables such an environment where different Metaverse apps can easily reside together and integrate. Moreover, efficiency gains via consolidation and infrastructure sharing is possible. 6G clouds can support ultra-scalable compute storage for spatiotemporal changes in the Metaverse services. However, the trade-off between latency and cloud centralization is an important research topic [133].

### F. 6G IOE
1) Introduction

The growth of IoT applications results in increasing the number of IoT devices, which is expected to grow up to 24 billion by 2030 [136]. Furthermore, the total IoT market will also grow up to USD 1.5 trillion in 2030. The dawn of Internet of Everything (IoE) is envisaged to expand the IoT paradigm to weave a hyper-connected network of not only things but also data, people, and processes [137]. Therefore, IoE is expected to integrate "Everything" for connecting, identifying, monitoring, and making intelligent decisions towards realizing new applications and services. IoE will connect many ecosystems involving heterogeneous sensors, actuators, user equipment, data types, services, and applications [138]. Numerous heterogeneous sensors in IoE can obtain data related to various parameters ranging from location, speed, acceleration, temperature, ambient light, humidity and air pressure to biosignals. This sensory information is paramount for the functionality of the Metaverse as real-world information provides inputs to form and update the virtual space and allow interactions between the real world and the virtual world. Furthermore, Human-Computer Interaction (HCI) can provide more flexible ways to access the Metaverse through human sensing (e.g. gesture recognition) [5]. Numerous cameras can capture video sequences from multiple angles to recognize human activities through advanced AI-enabled computer vision algorithms. In addition, the captured audio-visual information can be used to predict human emotions with the aid of smart wearables. These smart wearables can also capture data that are useful to obtain health metrics, such as heart rate, oxygen saturation level, body temperature, and electrocardiogram (ECG). 6G provides the ubiquitous, uninterruptible, ultra-high reliable/available and massive low-latency communication demanded by IoE [5], [137]. In addition, the edge-6G capabilities of 6G can process massive amounts of data collected from IoE devices to provide meaningful information for 6G applications and services. The integration of 6G and IoE will have the potential to enable many services, including the internet of medical things, smart healthcare, robotics, industry 5.0, smart grids, smart cities, and body area networks [137]. The superior connectivity offered through 6G with features such as, near real-time connectivity, extreme data rates, access to powerful computing resources at the network edge, and massive machine-type communication under strict delay constraints between heterogeneous sensory devices will facilitate the smooth operation of the Metaverse services and applications [139], [140].

### 2) How 6G IOE can help the Metaverse

6G IoE plays an important role towards enabling the Metaverse by supporting an extremely large number of users, sensors, and devices to connect and communicate seamlessly with extremely high data rates, ultra-low delays, and jitters [137]. In addition, the data obtained through heterogeneous IoE devices can be processed using AI and ML through powerful Multi-access Edge Computing (MEC) resources in envisaged 6G networks.

For instance, [141] discusses the expansion of IoE and how a multitude of sensors will enable the Extended Reality (XR) applications in the Metaverse. This work also explores the convergence of AI, MEC, Robots, and Distributed Ledger Technologies, such as blockchain, towards expanding the horizons of IoT towards IoE and beyond to provide a beyond smartphone experience. The proposed multisensory architecture is capable of integrating ubiquitous and pervasive computing towards enhancing human perception through advanced XR experiences. This is performed by utilizing wearables and nearby network resources in the 6G era. Hence, the dawn and the evolution of IoE will facilitate cross-reality environments, such as the Metaverse that can fuse real and virtual worlds with networked humans, avatars, and robots.

In addition, 6G IoE enables "wireless sensing" to sense the behavior of surrounding humans and the environment [5]. The functionality of IoT is expanded from simply networking a large number of devices towards sensing the wireless network. Various wireless signals including Wireless Fidelity (WiFi), Zigbee, Bluetooth, and Radio-Frequency IDentification (RFID) are used as sensing mediums through analyzing the signal variation (e.g. signal blocking, signal reflection, and signal scattering) caused by surrounding humans and objects [142]. These variations may change signal properties, such as phase, frequency and amplitude, which can be inferred through parameters including Received Signal Strength (RSS), Channel State Information (CSI), and Doppler shift. Together with signal preprocessing techniques, such as filtering and de-noising to minimize the effect of signal interference and noise, changes in the environment can be recognized by identifying distinguishable unique features owing to ML models. The accuracy of such predictions can be enhanced through the widespread of mmWave and MIMO technologies. In addition, an Integrated Sensing and Communication (ISAC) system, where communication systems and IoE hardware are jointly designed can improve the accuracy of wireless sensing while enhancing spectrum efficiency and minimizing hardware implementation cost [143]. However, modelling such systems, providing real-time access to powerful computational resources for data processing through advanced AI and ML schemes, and providing real-time ultra-low latency communication with seamless coverage requires beyond 5G network capabilities that are expected to be facilitated by emerging 6G networks.

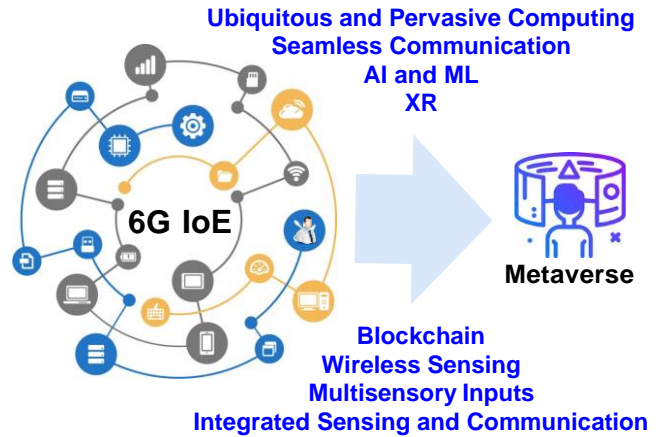

**FIGURE 13.** 6G IoE for the Metaverse

### 3) Summary

The evolution of IoT towards IoE with the dawn of 6G provides seamless connectivity, extreme data rates, ultra-low latency and ultra-high reliable/available communication, and real-time access to powerful Edge-AI-enabled computational resources to facilitate the Metaverse applications. 6G IoE also facilitates advanced wireless sensing with mmWave and MIMO technologies. The development of ISAC harnessing extreme communication capabilities and Edge-AI processing of 6G networks can further improve the capabilities of 6G IoE that would enable emerging the Metaverse applications. 6G IoE features that enable the Metaverse applications are illustrated in Fig. 13.

### G. OTHER 6G TECHNOLOGIES
### 1) Extended Reality

Extended Reality (XR) combines Virtual Reality (VR), Augmented Reality (AR) and Mixed Reality (MR) to blur the border between physical and virtual worlds with wearables supporting human-machine interactions with real and computer-generated environments [137]. 6G is capable of facilitating the massive low-latency, extremely low latency and extremely high data rate demanded by XR applications. Together with Edge-AI capabilities, 6G can facilitate the seamless 3C (computing, caching and communication) services for XR applications. Many sensors are used for the data collection on user location, orientation and movements. XR enables telepresence towards facilitating various aspects of human life, such as work, education, shopping, healthcare, tourism, and entertainment [144]. For instance, [145] explores how XR impacts six dimensions of workload, as defined by NASA Task Load Index (NASA-TLX), namely, mental demand, physical demand, temporal demand, performance, effort, and frustration, and the overall workload in the retail sector. The results of the study indicate that albeit VR alone did not have a significant impact on the various dimensions of workload, XR had a significant impact on performing shopping-related tasks. In addition, [146] presents how users can actively engage with 3D content to stimulate healthy behaviour using XR in the Metaverse. This work discusses how XR can be effectively used in the Metaverse to address long-term health issues and challenges. Accordingly, XR can be identified as an important enabler to provide services efficiently using the Metaverse. However, challenges, such as limitations in physical and cognitive resources, lack of experience with VR environments, and difficulties in using existing XR devices for prolonged periods, need to be addressed towards utilizing XR for the Metaverse applications in future 6G networks.

2) Digital Twins

The Metaverse applications demand next-generation networks to facilitate the high processing capabilities demanded by the Metaverse applications. These can be provided through the edge-AI capabilities of emerging 6G networks. Digital Twins (DT) can be an important enabler of the cloud-native network paradigm, which can efficiently support the Metaverse [147]. DTs act as a digital representation of humans and things in cyberspace. Cybertwins can provide a multitude of services for the Metaverse, including, acting as a communication assistant, logging network behavior, and own digital assets, in a flexible, scalable, secure and reliable manner. 6G IoE can play a key role towards facilitating DTs. In [148], the authors discuss how to utilize a cloud network operating system that can work distributively in a real-time multi-agent platform to allocate 3C resources, which are considered to be integral components of envisaged 6G networks [137]. In addition, the Metaverse applications demand 6G networks to support intelligent and fully autonomous operation. In response [149] proposes a Digital Twin Edge Network (DITEN). DITEN is able to combine Multi-access Edge Computing (MEC) together with DT to improve the network throughput, enhance network security, and reduce the cost of 3C services. DITEN continuously monitors the network status for DT modelling, updating and deployment and performs tasks such as routing and resource management efficiently to enable applications such as the Metaverse. However, there are several open issues and challenges, including high-precision DT modelling, DT migration for mobility and ensuring security and privacy.

3) Space-Air-Ground Integrated Network (SAGIN)

Global sensing and seamless connectivity are paramount to providing uninterrupted access to the Metaverse applications through 6G networks. However, ground networks alone are not capable of providing ubiquitous connectivity to the Metaverse applications in a reliable and cost-efficient fashion [149]. This is even evident in mountain areas and in disastrous situations. As a solution, Non-Terrestrial Networks (NTN) Towards 3D Networking are proposed with 6G networks [137]. NTN provides 3D network coverage and backhauling through integrating Unmanned Aerial Vehicles (UAVs), satellites, balloons and High Altitude Platform (HAP) stations [150]. 3D networking expands the NTN paradigm through incorporating space, underground, and underwater communication [151]. For instance, project 3GPP TR 38.811 intends to support non-terrestrial networks by considering the architecture and channel models across satellite, air access, and terrestrial cellular networks [137]. In addition, multi-dimensional networks named Space-Air-Ground Integrated Network (SAGIN) envisage to deeply integrate of space nodes (e.g. satellites), air nodes (e.g. UAVs, drones, air balloons), and terrestrial network nodes (e.g. 5G and beyond network nodes) towards providing seamless connectivity [5]. However, the seamless inter-operation and resource management among multiple types of networks require unified access methods and network standards towards facilitating seamless connectivity for the Metaverse applications.

## V. 6G INTEGRATION CHALLENGES

In this section, we present the challenges raised by limited backwards compatibility with existing devices, lack of standards, accountability, resilience & privacy preservation, energy inefficiency, and radio design & carrier bandwidths while integrating 6G with the Metaverse.

### A. LIMITED BACKWARDS COMPATIBILITY WITH EXISTING DEVICES

1) Introduction to issues

Effective communication in the Metaverse requires compatibility with previous-generation networks such as 4G and 5G. Despite that, some Metaverse applications can operate on existing network capabilities devices due to the deployment of 6G these devices become worthless.

2) Possible solutions

A potential solution to address this issue is the backward compatibility of the 6G network with existing devices that enables the addition of high-capacity communication in the Metaverse and also delivers faster data rates for applications requiring real-time processing and integration. The 6G networks should support the features of the previous generations of communications like the 5G network for some time, enabling progressive migration of the Metaverse devices and lowering the overall cost of 6G and the Metaverse integration. In order to evaluate backward compatibility, mobile operators need to consider how the 5G and 6G core networks are connected and work on the 3GPP standard accordingly.

### B. LACK OF STANDARDS

1) Introduction to issues

There is a concern among users about the Metaverse's potential legal consequences. If a problem arises, there is no agreed-upon policy framework or set of standards for the integration of 6G with the Metaverse. Any problem with the integration of these technologies will affect the trust and the capabilities of the 6G networks and the Metaverse.

**TABLE 4.** Summary of related works

| | 6G for the Metaverse - technical perspective | | | | | | | |
|---|---|---|---|---|---|---|---|---|
| Ref. | Validation of digital assets | Cross platform integration and interoperability | Efficient support of AI | High speed data connection | Low Latency Communication | Computer Vision | High transaction integration | Security and privacy |
| 7 | | | | | | | | x |
| 30-34 | x | | | | | x | | |
| 35-38 | | x | | | | | | x |
| 39-41 | | | x | | | | | |
| 42-47 | | | | x | x | x | | |
| 48-49 | | | | x | x | x | | |
| 50-53 | | | x | | | x | | |
| 54-56 | | | | | | | x | |
| | The role of 6G technologies for the Metaverse | | | | | | | |
| Ref. | AI | Blockchain | Edge | OpenRAN | Cloud | IoE/IoT | XR | Digital Twin |
| 57-84 | x | | | | | | x | |
| 85-104 | | x | x | | | | | |
| 105-125 | x | | x | x | x | | | |
| 126-130 | | | | x | | | | |
| 131-134 | | | | | x | | | |
| 135-142 | | x | | | x | x | x | |
| 2, 143-148 | x | x | x | | x | x | x | x |

### 2) Possible solutions

These challenges may be resolved by establishing a forum involving service providers, researchers, and legal counsel to develop standards and policy frameworks that address concerns about user ethics, safety, and privacy while integrating 6G with the Metaverse. The users should be provided with complete control and transparency of their data transmitted over 6G networks, which ensures their privacy in the Metaverse. As a consequence, this will raise the bar for the 6G communication networks and the Metaverse, which will increase trust among the users. For example, though ORAN is not yet fully functional it has an alliance focusing on the integration issues of multiple service providers which will enhance the bandwidth availability and security of the overall networks.

### C. ACCOUNTABILITY, RESILIENCE AND PRIVACY PRESERVATION

1) Introduction to issues

The functionalities across 6G integrated Metaverse will be mostly automated based on the decisions made by AI. Any misclassification made by these decisions that cannot be traced because of the black box nature of AI will have a direct effect on the accountability of the 6G integrated Metaverse.

2) Possible solutions

Explainable AI (XAI) is a promising solution for this issue which allows us to understand the misclassification issues and improve trust in the decisions made in the 6G integrated Metaverse. The usage of xAI will aid in pinpointing the problem's cause, assist the Metaverse's administrators in understanding the issue, and motivate them to prevent a recurrence - this enhances the transparency of auditing of issues related to the 6G integrated Metaverse. Additionally, existing and newly proposed AI algorithms need to be analysed considering their accountability, resilience and privacy preservation capabilities within the context of future networks.

### D. ENERGY INEFFICIENCY

1) Introduction to issues

The integration of processing, communication, sensing, and control capabilities inside a 6G network enables a seamless transition between the virtual and physical worlds, consequently contributing to the realisation of the Metaverse. To support the requirements of the Metaverse, the cellular capacity should be increased on top of the existing network infrastructure. This will require 6G to deploy more microscopic and even micro-cells in the network. This increases technological and network complexity and will further strain the energy efficiency and sustainability of the Metaverse.

2) Possible solutions

The integration of AI with 6G will address the issues of energy efficiency and network complexity, opening the door to a sustainable Metaverse ecosystem. The use of Zero touch network & Service Management (ZSM) in 6G provides an intelligent network for the Metaverse by enabling effective data access and cross-domain data exposure by permitting operational data to be maintained apart from the management applications. This will also improve the reliability of communication in the Metaverse.

### E. RADIO DESIGN AND CARRIER BANDWIDTHS

1) Introduction to issues

One of the main goals of 6G is to achieve Tb/s data rates, which requires large bandwidths (10-100 GHz spectrum for THz bands), which requires an aggregation of a large number of carriers to create larger bandwidth. Designing radios that work at sub-THz bands present a significant challenge to the industry and research due to the complexity of associated RF circuits. Finding the right balance in terms of transceiver efficiency, power generation, heat dissipation and the cost is critical for the successful adoption of radios to sub-THz bands.

2) Possible solutions

6G should provide more bandwidth and lower latency to improve the overall connectivity of the Metaverse. On 6G networks, there should be a 10 to 100-fold reduction in latency and an increase in bandwidth capacity for the users of the Metaverse to have the best immersive experiences. Every piece of networking hardware must have its material, component manufacture, and antenna architecture modified. To comply with the 6G standard, base station operations must change. 6G should depend on tightly focused, packaged radio signals rather than "omnidirectional" radio channels. Moreover, tightly focused radio signals use less energy, have high transceiver efficiency, less heat dissipation and less cost.

## VI. 6G METAVERSE PROJECTS

This section provides an overview of research projects and developments that are already underway towards realizing the Metaverse by harnessing the extreme network capabilities of envisioned B5G and 6G mobile networks.

### A. META

Meta, formerly known as Facebook, is presently working on combining social media with VR and AR towards realizing the Metaverse for users to work, play and interact with other users online [152]. This is possible due to the extreme mobile broadband capabilities, near zero latency, extreme reliability and availability, and network intelligence of emerging mobile networks. Users can join the Metaverse using VR headsets. The envisaged applications will range from connecting people, education, training, healthcare and the workplace to gaming. For instance, education technologies are expected to broaden their horizons from platforms to passively absorb information to learn by doing and experiencing through 3D immersion. In addition, Meta is working on building the Metaverse responsibly ensuring a safe, secure, and transparent operation. Meta has also launched the Meta Immersive Learning Academy and Research Fund to collaborate in building a solid and interoperable Metaverse platform. In addition, their Spark AR platform enables the creation and sharing of AR experiences through their apps and devices. Furthermore, Meta is working on building economic opportunities in the Metaverse to maintain and thrive in a digital economy in the future.

### B. VR HIVE

VR Hive [153] aims to transform e-learning through VR from the comfort of home or workplace. This project aims to design and develop a fully immersive learning platform over 6G mobile networks to feature the Metaverse that can be used to provide education, training, holographic telepresence, and real-time communication. These features will be provided through the extreme network capabilities of emerging 6G networks, such as, near real-time ultra-reliable communication with ultra-low latency and edge intelligence. Relevant infrastructure and network-aware immersive and adaptive environments will be developed to facilitate education through the range of products offered through VR Hive.

### C. 6G LIFE

6G Life [154] aims to facilitate the envisaged digital transformation where 6G mobile networks will play a significant role in this revolution. The project not only aims to develop the digital infrastructure and high-performance computing platforms but also concentrates on political and social issues that are required to realize future 6G applications. Realizing 6G applications will require diverse communication capabilities including human-machine interaction in virtual worlds, such as the Metaverse. The project aims to provide innovative solutions in the areas of scalable communication, flexible software concepts, and adaptive hardware platforms. The four key aspects considered by the project are latency, resilience, security and sustainability. The research work, including both basic and applied research, is mainly performed considering Industry 4.0/5.0, and intelligent healthcare applications.

### D. DECENTRALAND

Decentraland [155] is a decentralized virtual world where users can create objects, trade and interact with others in a virtual world. This also allows users to control policies on the operation of the virtual world. Decentraland operates as a Decentralized Autonomous Organization (DAO), where it owns smart contracts and assets on virtual land and estate contracts, wearables and other devices, and the marketplace to trade virtual assets. These developments can be realized through the capabilities of emerging 6G mobile networks, where extreme mobile connectivity will facilitate seamless connectivity to the virtual world. Furthermore, blockchain operation and smart contract execution will be enabled through the edge computing capabilities of the 6G networks.

Similar projects, such as Sandbox [156], Axie Infinity [157], and Illuvium [158] also envisage harnessing the capabilities of blockchain and emerging mobile networks towards realizing the Metaverse.

### E. LUXEMBOURG METAVERSE

The Luxembourg Metaverse [159] project aims to build a digital twin of an area of Luxembourg City. These digital twins can be explored by the public and the industry to provide multiple working opportunities. Luxemburg 5G-6G network digital twin aims to enable seamless and highly capable network connectivity to facilitate real-time services banking on emerging communication networks, such as beyond 5G and 6G. This project will also raise awareness of the advantages and applications of the Metaverse to the public and the industry. Furthermore, the project expects to optimise and secure the Metaverse deployments while integrating the latest developments of networks in a cost-effective and cost-efficient manner.

TABLE 5. 6G Metaverse Projects

| Project | Objective | 6G Technologies | | | | | | | |
|---|---|---|---|---|---|---|---|---|---|
| | | AI | Blockchain | Edge | OpenRAN | Cloud | IoE | XR | Digital Twin |
| Meta | Combine social media with VR and AR to facilitate work, play and other interactions among users online | ✓ | ✓ | ✓ | ✓ | ✓ | | ✓ | ✓ |
| VR Hive | Transform e-learning through VR to be accessed at home or workplace | ✓ | | | | ✓ | | ✓ | |
| 6G Life | Facilitate the digital transformation towards 6G with human machine collaboration | ✓ | | ✓ | ✓ | ✓ | ✓ | | ✓ |
| Decentraland | Create a virtual world for users to create objects, trade and interact with users | ✓ | ✓ | ✓ | | ✓ | ✓ | ✓ | ✓ |
| Luxembourg Metaverse | Build a digital twin of an area of Luxembourg city | ✓ | | | | ✓ | ✓ | ✓ | ✓ |

The 6G technological directions explored by the 6G metaverse projects presented in this section are tabulated in TABLE 5.

## VII. CONCLUSION

This paper presents the role of 6G towards realizing the Metaverse applications and services. The paper presents the role of 6G technologies in the immersive, smart, scalable and secure realization of the Metaverse. Furthermore, the paper presents how various 6G capabilities play a key role towards the realization of the Metaverse, including the role of 6G for, cross-platform integration, efficient support for AI, high-speed data connectivity, efficient user interaction, low latency communication, computer vision, high transaction integration, and security and privacy protection. Consequently, the integration challenges of 6G with the Metaverse are elaborated while providing several research directions towards realizing the Metaverse owing to the capabilities of future 6G networks.

## REFERENCES

[1] L.-H. Lee, T. Braud, P. Zhou, L. Wang, D. Xu, Z. Lin, A. Kumar, C. Bermejo, and P. Hui, "All one needs to know about metaverse: A complete survey on technological singularity, virtual ecosystem, and research agenda," arXiv preprint arXiv:2110.05352, 2021.
[2] "Cloud vr network solution white paper."
[3] T. Huynh-The, Q.-V. Pham, X.-Q. Pham, T. T. Nguyen, Z. Han, and D.-S. Kim, "Artificial intelligence for the metaverse: A survey," arXiv preprint arXiv:2202.10336, 2022.
[4] Y. Wang, Z. Su, N. Zhang, R. Xing, D. Liu, T. H. Luan, and X. Shen, "A survey on metaverse: Fundamentals, security, and privacy," IEEE Communications Surveys & Tutorials, pp. 1–1, 2022.
[5] F. Tang, X. Chen, M. Zhao, and N. Kato, "The roadmap of communication and networking in 6g for the metaverse," IEEE Wireless Communications, 2022.
[6] M. A. I. Mozumder, M. M. Sheeraz, A. Athar, S. Aich, and H.-C. Kim, "Overview: technology roadmap of the future trend of metaverse based on iot, blockchain, ai technique, and medical domain metaverse activity," in 2022 24th International Conference on Advanced Communication Technology (ICACT), pp. 256–261, IEEE, 2022.
[7] M. Xu, W. C. Ng, W. Y. B. Lim, J. Kang, Z. Xiong, D. Niyato, Q. Yang, X. Sherman Shen, and C. Miao, "A Full Dive into Realizing the Edge-enabled Metaverse: Visions, Enabling Technologies,and Challenges," arXiv e-prints, p. arXiv:2203.05471, Mar. 2022.
[8] L. Chang, Z. Zhang, P. Li, S. Xi, W. Guo, Y. Shen, Z. Xiong, J. Kang, D. Niyato, X. Qiao, et al., "6g-enabled edge ai for metaverse: Challenges, methods, and future research directions," arXiv preprint arXiv:2204.06192, 2022.
[9] C. D. Alwis, A. Kalla, Q.-V. Pham, P. Kumar, K. Dev, W.-J. Hwang, and M. Liyanage, "Survey on 6g frontiers: Trends, applications, requirements, technologies and future research," IEEE Open Journal of the Communications Society, vol. 2, pp. 836–886, 2021.
[10] T. S. Rappaport, Y. Xing, O. Kanhere, S. Ju, A. Madanayake, S. Mandal, A. Alkhateeb, and G. C. Trichopoulos, "Wireless communications and applications above 100 ghz: Opportunities and challenges for 6g and beyond," IEEE access, vol. 7, pp. 78729–78757, 2019.
[11] A. Salh, L. Audah, N. S. M. Shah, A. Alhammadi, Q. Abdullah, Y. H. Kim, S. A. Al-Gailani, S. A. Hamzah, B. A. F. Esmail, and A. A. Almohammedi, "A survey on deep learning for ultra-reliable and low-latency communications challenges on 6g wireless systems," IEEE Access, vol. 9, pp. 55098–55131, 2021.
[12] X. Shen, W. Liao, and Q. Yin, "A novel wireless resource management for the 6g-enabled high-density internet of things," IEEE Wireless Communications, vol. 29, no. 1, pp. 32–39, 2022.
[13] Y. Chen, W. Liu, Z. Niu, Z. Feng, Q. Hu, and T. Jiang, "Pervasive intelligent endogenous 6g wireless systems: Prospects, theories and key technologies," Digital communications and networks, vol. 6, no. 3, pp. 312–320, 2020.
[14] X. Shen, W. Liao, and Q. Yin, "A novel wireless resource management for the 6g-enabled high-density internet of things," IEEE Wireless Communications, vol. 29, no. 1, pp. 32–39, 2022.
[15] S. Kraus, D. K. Kanbach, P. M. Krysta, M. M. Steinhoff, and N. Tomini, "Facebook and the creation of the metaverse: radical business model innovation or incremental transformation?," International Journal of Entrepreneurial Behavior & Research, 2022.
[16] S. Hollensen, P. Kotler, and M. O. Opresnik, "Metaverse–the new marketing universe," Journal of Business Strategy, 2022.
[17] H. R. Hasan and K. Salah, "Proof of delivery of digital assets using blockchain and smart contracts," IEEE Access, vol. 6, pp. 65439–65448, 2018.
[18] Å. Fast-Berglund, L. Gong, and D. Li, "Testing and validating extended reality (xr) technologies in manufacturing," Procedia Manufacturing, vol. 25, pp. 31–38, 2018.
[19] A. Siyaev and G.-S. Jo, "Neuro-symbolic speech understanding in aircraft maintenance metaverse," IEEE Access, vol. 9, pp. 154484–154499, 2021.
[20] T. Zhang, L. Gao, L. He, M. Zhang, B. Krishnamachari, and A. S. Avestimehr, "Federated learning for the internet of things: Applications, challenges, and opportunities," IEEE Internet of Things Magazine, vol. 5, no. 1, pp. 24–29, 2022.
[21] Y. Lu, X. Huang, K. Zhang, S. Maharjan, and Y. Zhang, "Communication-efficient federated learning for digital twin edge networks in industrial iot," IEEE Transactions on Industrial Informatics, vol. 17, no. 8, pp. 5709–5718, 2020.

[22] I. Skalidis, O. Muller, and S. Fournier, "Cardioverse: The cardiovascular medicine in the era of metaverse," Trends in Cardiovascular Medicine, 2022.

[23] Surfing the Metaverse's Real Estate Boom.

[24] A. Beniiche, S. Rostami, and M. Maier, "Society 5.0: Internet as if people mattered," IEEE Wireless Communications, 2022.

[25] D. Gursoy, S. Malodia, and A. Dhir, "The metaverse in the hospitality and tourism industry: An overview of current trends and future research directions," Journal of Hospitality Marketing & Management, pp. 1–8, 2022.

[26] D. Shin, "The actualization of meta affordances: Conceptualizing affordance actualization in the metaverse games," Computers in Human Behavior, vol. 133, p. 107292, 2022.

[27] I. A. Ilyina, E. A. Eltikova, K. A. Uvarova, and S. D. Chelysheva, "Metaverse-death to offline communication or empowerment of interaction?," in 2022 Communication Strategies in Digital Society Seminar (ComSDS), pp. 117–119, IEEE, 2022.

[28] B. Falchuk, S. Loeb, and R. Neff, "The social metaverse: Battle for privacy," IEEE Technology and Society Magazine, vol. 37, no. 2, pp. 52–61, 2018.

[29] T. R. Gadekallu, T. Huynh-The, W. Wang, G. Yenduri, P. Ranaweera, Q.-V. Pham, D. B. da Costa, and M. Liyanage, "Blockchain for the metaverse: A review," arXiv preprint arXiv:2203.09738, 2022.

[30] J. Kim, "Advertising in the metaverse: Research agenda," Journal of Interactive Advertising, vol. 21, no. 3, pp. 141–144, 2021.

[31] M. Nadini, L. Alessandretti, F. Di Giacinto, M. Martino, L. M. Aiello, and A. Baronchelli, "Mapping the nft revolution: market trends, trade networks, and visual features," Scientific reports, vol. 11, no. 1, pp. 1–11, 2021.

[32] H. Ko, B. Son, Y. Lee, H. Jang, and J. Lee, "The economic value of nft: Evidence from a portfolio analysis using mean–variance framework," Finance Research Letters, vol. 47, p. 102784, 2022.

[33] A. Park, J. Kietzmann, L. Pitt, and A. Dabirian, "The evolution of nonfungible tokens: Complexity and novelty of nft use-cases," IT Professional, vol. 24, no. 1, pp. 9–14, 2022.

[34] D. Vidal-Tomás, "The new crypto niche: Nfts, play-to-earn, and metaverse tokens," Finance Research Letters, p. 102742, 2022.

[35] C. Hackl, D. Lueth, and T. Di Bartolo, Navigating the Metaverse: A Guide to Limitless Possibilities in a Web 3.0 World. John Wiley & Sons, 2022.

[36] E. Lee, Y.-D. Seo, S.-R. Oh, and Y.-G. Kim, "A survey on standards for interoperability and security in the internet of things," IEEE Communications Surveys & Tutorials, vol. 23, no. 2, pp. 1020–1047, 2021.

[37] A. Hazra, M. Adhikari, T. Amgoth, and S. N. Srirama, "A comprehensive survey on interoperability for iiot: taxonomy, standards, and future directions," ACM Computing Surveys (CSUR), vol. 55, no. 1, pp. 1–35, 2021.

[38] M. Sparkes, "What is a metaverse," 2021.

[39] S. A. Khan, I. Shayea, M. Ergen, and H. Mohamad, "Handover management over dual connectivity in 5g technology with future ultra-dense mobile heterogeneous networks: A review," Engineering Science and Technology, an International Journal, p. 101172, 2022.

[40] S. Mystakidis, "Metaverse," Encyclopedia, vol. 2, no. 1, pp. 486–497, 2022.

[41] M. Campbell and M. Jovanovic, "Digital self: The next evolution of the digital human," Computer, vol. 55, no. 04, pp. 82–86, 2022.

[42] B. K. Wiederhold, "Ready (or not) player one: initial musings on the metaverse," 2022.

[43] H. Duan, J. Li, S. Fan, Z. Lin, X. Wu, and W. Cai, "Metaverse for social good: A university campus prototype," in Proceedings of the 29th ACM International Conference on Multimedia, pp. 153–161, 2021.

[44] M. Inomata, W. Yamada, N. Kuno, M. Sasaki, K. Kitao, M. Nakamura, H. Ishikawa, and Y. Oda, "Terahertz propagation characteristics for 6g mobile communication systems," in 2021 15th European Conference on Antennas and Propagation (EuCAP), pp. 1–5, IEEE, 2021.

[45] D.-I. D. Han, Y. Bergs, and N. Moorhouse, "Virtual reality consumer experience escapes: preparing for the metaverse," Virtual Reality, pp. 1–16, 2022.

[46] C. Girvan, "What is a virtual world? definition and classification," Educational Technology Research and Development, vol. 66, no. 5, pp. 1087–1100, 2018.

[47] D. M. Markowitz, R. Laha, B. P. Perone, R. D. Pea, and J. N. Bailenson, "Immersive virtual reality field trips facilitate learning about climate change," Frontiers in psychology, vol. 9, p. 2364, 2018.

[48] D. Mendes, F. M. Caputo, A. Giachetti, A. Ferreira, and J. Jorge, "A survey on 3d virtual object manipulation: From the desktop to immersive virtual environments," in Computer graphics forum, vol. 38, pp. 21–45, Wiley Online Library, 2019.

[49] R. Suzuki, A. Karim, T. Xia, H. Hedayati, and N. Marquardt, "Augmented reality and robotics: A survey and taxonomy for ar-enhanced human-robot interaction and robotic interfaces," in CHI Conference on Human Factors in Computing Systems, pp. 1–33, 2022.

[50] M. Adhikari and A. Hazra, "6g-enabled ultra-reliable low-latency communication in edge networks," IEEE Communications Standards Magazine, vol. 6, no. 1, pp. 67–74, 2022.

[51] M.-H. Guo, T.-X. Xu, J.-J. Liu, Z.-N. Liu, P.-T. Jiang, T.-J. Mu, S.-H. Zhang, R. R. Martin, M.-M. Cheng, and S.-M. Hu, "Attention mechanisms in computer vision: A survey," Computational Visual Media, pp. 1–38, 2022.

[52] T. Braud, C. B. Fernández, and P. Hui, "Scaling-up ar: University campus as a physical-digital metaverse," in 2022 IEEE Conference on Virtual Reality and 3D User Interfaces Abstracts and Workshops (VRW), pp. 169–175, IEEE, 2022.

[53] H. Kumar et al., "Epidemiological mucormycosis treatment and diagnosis challenges using the adaptive properties of computer vision techniques based approach: a review," Multimedia Tools and Applications, pp. 1–29, 2022.

[54] C. Yeh, G. Do Jo, Y.-J. Ko, and H. K. Chung, "Perspectives on 6g wireless communications," ICT Express, 2022.

[55] Y. Liu, F. R. Yu, X. Li, H. Ji, and V. C. Leung, "Decentralized resource allocation for video transcoding and delivery in blockchain-based system with mobile edge computing," IEEE Transactions on Vehicular Technology, vol. 68, no. 11, pp. 11169–11185, 2019.

[56] D. Frey, J. Royan, R. Piegay, A.-M. Kermarrec, E. Anceaume, and F. Le Fessant, "Solipsis: A decentralized architecture for virtual environments," in 1st International Workshop on Massively Multiuser Virtual Environments, 2008.

[57] C. Nguyen, D. Hoang, D. Nguyen, and E. Dutkiewicz, "Metachain: a novel blockchain-based framework for metaverse applications. arxiv," arXiv preprint arXiv:2201.00759, 2021.

[58] T. Huynh-The, Q.-V. Pham, X.-Q. Pham, T. T. Nguyen, Z. Han, and D.-S. Kim, "Artificial intelligence for the metaverse: A survey," arXiv preprint arXiv:2202.10336, 2022.

[59] A. Fuad and M. Al-Yahya, "Recent developments in arabic conversational ai: A literature review," IEEE Access, vol. 10, pp. 23842–23859, 2022.

[60] P.-S. Chiu, J.-W. Chang, M.-C. Lee, C.-H. Chen, and D.-S. Lee, "Enabling intelligent environment by the design of emotionally aware virtual assistant: A case of smart campus," IEEE Access, vol. 8, pp. 62032–62041, 2020.

[61] F. Cui, Q. Cui, and Y. Song, "A survey on learning-based approaches for modeling and classification of human–machine dialog systems," IEEE Transactions on Neural Networks and Learning Systems, vol. 32, no. 4, pp. 1418–1432, 2021.

[62] I. Giorgi, B. Golosio, M. Esposito, A. Cangelosi, and G. L. Masala, "Modeling multiple language learning in a developmental cognitive architecture," IEEE Transactions on Cognitive and Developmental Systems, vol. 13, no. 4, pp. 922–933, 2021.

[63] D. W. Otter, J. R. Medina, and J. K. Kalita, "A survey of the usages of deep learning for natural language processing," IEEE Transactions on Neural Networks and Learning Systems, vol. 32, no. 2, pp. 604–624, 2021.

[64] G. Xu, Y. Meng, X. Qiu, Z. Yu, and X. Wu, "Sentiment analysis of comment texts based on bilstm," IEEE Access, vol. 7, pp. 51522–51532, 2019.

[65] Y. Cheng, L. Yao, G. Xiang, G. Zhang, T. Tang, and L. Zhong, "Text sentiment orientation analysis based on multi-channel cnn and bidirectional gru with attention mechanism," IEEE Access, vol. 8, pp. 134964–134975, 2020.

[66] Z. Hu, A. Bulling, S. Li, and G. Wang, "Fixationnet: Forecasting eye fixations in task-oriented virtual environments," IEEE Transactions on Visualization and Computer Graphics, vol. 27, pp. 2681–2690, Mar. 2021.

[67] R. Miller, N. K. Banerjee, and S. Banerjee, "Combining real-world constraints on user behavior with deep neural networks for virtual reality (vr) biometrics," in 2022 IEEE Conference on Virtual Reality and 3D User Interfaces (VR), (Christchurch, New Zealand), pp. 409–418, 2022.

[68] Y. Jin, M. Chen, T. Goodall, A. Patney, and A. C. Bovik, "Subjective and objective quality assessment of 2d and 3d foveated video compression in virtual reality," IEEE Transactions on Image Processing, vol. 30, pp. 5905–5919, Jun. 2021.

[69] T. Huynh-The, C.-H. Hua, N. A. Tu, T. Hur, J. Bang, D. Kim, M. B. Amin, B. H. Kang, H. Seung, S.-Y. Shin, et al., "Hierarchical topic modeling with pose-transition feature for action recognition using 3d skeleton data," Information Sciences, vol. 444, pp. 20–35, May 2018.

[70] T. Huynh-The, C.-H. Hua, N. A. Tu, and D.-S. Kim, "Learning 3d spatiotemporal gait feature by convolutional network for person identification," Neurocomputing, vol. 397, pp. 192–202, Jul. 2020.

[71] T. Huynh-The, C.-H. Hua, and D.-S. Kim, "Encoding pose features to images with data augmentation for 3-d action recognition," IEEE Transactions on Industrial Informatics, vol. 16, pp. 3100–3111, May 2020.

[72] C.-H. Hua, T. Huynh-The, and S. Lee, "Convolutional networks with bracket-style decoder for semantic scene segmentation," in Proc. 2018 IEEE International Conference on Systems, Man, and Cybernetics (SMC), (Miyazaki, Japan), pp. 2980–2985, IEEE, 2018.

[73] C.-H. Hua, T. Huynh-The, S.-H. Bae, and S. Lee, "Cross-attentional bracket-shaped convolutional network for semantic image segmentation," Information Sciences, vol. 539, pp. 277–294, Oct. 2020.

[74] L.-C. Chen, G. Papandreou, I. Kokkinos, K. Murphy, and A. L. Yuille, "Deeplab: Semantic image segmentation with deep convolutional nets, atrous convolution, and fully connected crfs," IEEE Transactions on Pattern Analysis and Machine Intelligence, vol. 40, pp. 834–848, Apr. 2018.

[75] J. Kim, H. Zeng, D. Ghadiyaram, S. Lee, L. Zhang, and A. C. Bovik, "Deep convolutional neural models for picture-quality prediction: Challenges and solutions to data-driven image quality assessment," IEEE Signal Processing Magazine, vol. 34, pp. 130–141, Nov. 2017.

[76] R. Ranjan, V. M. Patel, and R. Chellappa, "Hyperface: A deep multi-task learning framework for face detection, landmark localization, pose estimation, and gender recognition," IEEE Transactions on Pattern Analysis and Machine Intelligence, vol. 41, pp. 121–135, Dec. 2019.

[77] X. Wang, L. Gao, J. Song, and H. Shen, "Beyond frame-level cnn: Saliency-aware 3-d cnn with lstm for video action recognition," IEEE Signal Processing Letters, vol. 24, pp. 510–514, Sep. 2017.

[78] Y. Xin, L. Kong, Z. Liu, Y. Chen, Y. Li, H. Zhu, M. Gao, H. Hou, and C. Wang, "Machine learning and deep learning methods for cybersecurity," IEEE Access, vol. 6, pp. 35365–35381, 2018.

[79] R. Vinayakumar, M. Alazab, K. P. Soman, P. Poornachandran, A. Al-Nemrat, and S. Venkatraman, "Deep learning approach for intelligent intrusion detection system," IEEE Access, vol. 7, pp. 41525–41550, 2019.

[80] A. Jamalipour and S. Murali, "A taxonomy of machine-learning-based intrusion detection systems for the internet of things: A survey," IEEE Internet of Things Journal, vol. 9, no. 12, pp. 9444–9466, 2022.

[81] Y. Liu, J. Wang, J. Li, S. Niu, L. Wu, and H. Song, "Zero-bias deep-learning-enabled quickest abnormal event detection in iot," IEEE Internet of Things Journal, vol. 9, pp. 11385–11395, Jul. 2022.

[82] Z. Lv, L. Qiao, J. Li, and H. Song, "Deep-learning-enabled security issues in the internet of things," IEEE Internet of Things Journal, vol. 8, pp. 9531–9538, Jun. 2021.

[83] A. Uprety and D. B. Rawat, "Reinforcement learning for iot security: A comprehensive survey," IEEE Internet of Things Journal, vol. 8, pp. 8693–8706, Jun. 2021.

[84] X. Liu, H. Li, G. Xu, S. Liu, Z. Liu, and R. Lu, "Padl: Privacy-aware and asynchronous deep learning for iot applications," IEEE Internet of Things Journal, vol. 7, pp. 6955–6969, Aug. 2020.

[85] W. Zhang, B. Jiang, M. Li, and X. Lin, "Privacy-preserving aggregate mobility data release: An information-theoretic deep reinforcement learning approach," IEEE Transactions on Information Forensics and Security, vol. 17, pp. 849–864, 2022.

[86] N. Deepa, Q.-V. Pham, D. C. Nguyen, S. Bhattacharya, B. Prabadevi, T. R. Gadekallu, P. K. R. Maddikunta, F. Fang, and P. N. Pathirana, "A survey on blockchain for big data: Approaches, opportunities, and future directions," Future Generation Computer Systems, vol. 131, pp. 209–226, Jun. 2022.

[87] C. T. Nguyen, D. T. Hoang, D. N. Nguyen, and E. Dutkiewicz, "Metachain: A novel blockchain-based framework for metaverse applications," arXiv preprint arXiv:2201.00759, 2022.

[88] K. Cheng, S. Quan, and J. Yan, "A blockchain-based crowdsourcing system for large scale environmental data acquisition," in 2021 IEEE 24th International Conference on Computer Supported Cooperative Work in Design (CSCWD), (Dailan, China), pp. 855–860, May 2021.

[89] C. H. Liu, Q. Lin, and S. Wen, "Blockchain-enabled data collection and sharing for industrial iot with deep reinforcement learning," IEEE Transactions on Industrial Informatics, vol. 15, pp. 3516–3526, Jun. 2019.

[90] M. Li, J. Weng, A. Yang, W. Lu, Y. Zhang, L. Hou, J.-N. Liu, Y. Xiang, and R. H. Deng, "Crowdbc: A blockchain-based decentralized framework for crowdsourcing," IEEE Transactions on Parallel and Distributed Systems, vol. 30, pp. 1251–1266, Jun. 2019.

[91] U. Bodkhe, S. Tanwar, K. Parekh, P. Khanpara, S. Tyagi, N. Kumar, and M. Alazab, "Blockchain for industry 4.0: A comprehensive review," IEEE Access, vol. 8, pp. 79764–79800, 2020.

[92] R. Li, T. Song, B. Mei, H. Li, X. Cheng, and L. Sun, "Blockchain for large-scale internet of things data storage and protection," IEEE Transactions on Services Computing, vol. 12, pp. 762–771, Sept.-Oct. 2019.

[93] H. Shafagh, L. Burkhalter, A. Hithnawi, and S. Duquennoy, "Towards blockchain-based auditable storage and sharing of iot data," in Proceedings of the 2017 on Cloud Computing Security Workshop, (Dallas, Texas, USA), p. 45–50, Nov. 2017.

[94] W. Liang, Y. Fan, K.-C. Li, D. Zhang, and J.-L. Gaudiot, "Secure data storage and recovery in industrial blockchain network environments," IEEE Transactions on Industrial Informatics, vol. 16, pp. 6543–6552, Oct. 2020.

[95] R. Jabbar, N. Fetais, M. Krichen, and K. Barkaoui, "Blockchain technology for healthcare: Enhancing shared electronic health record interoperability and integrity," in 2020 IEEE International Conference on Informatics, IoT, and Enabling Technologies (ICIoT), (Doha, Qatar), pp. 310–317, Feb. 2020.

[96] W. J. Gordon and C. Catalini, "Blockchain technology for healthcare: Facilitating the transition to patient-driven interoperability," Computational and Structural Biotechnology Journal, vol. 16, pp. 224–230, Jul. 2018.

[97] X. Liu, S. X. Sun, and G. Huang, "Decentralized services computing paradigm for blockchain-based data governance: Programmability, interoperability, and intelligence," IEEE Transactions on Services Computing, vol. 13, pp. 343–355, Mar.-Apr. 2020.

[98] S. Schulte, M. Sigwart, P. Frauenthaler, and M. Borkowski, "Towards blockchain interoperability," in Proc. International Conference on Business Process Management, (Vienna, Austria), pp. 3–10, Springer, 2019.

[99] Y. Jiang, C. Wang, Y. Wang, and L. Gao, "A cross-chain solution to integrating multiple blockchains for iot data management," Sensors, vol. 19, May 2019.

[100] B. Pillai, K. Biswas, and V. Muthukkumarasamy, "Cross-chain interoperability among blockchain-based systems using transactions," The Knowledge Engineering Review, vol. 35, p. e23, Jun. 2020.

[101] P. Wei, D. Wang, Y. Zhao, S. K. S. Tyagi, and N. Kumar, "Blockchain data-based cloud data integrity protection mechanism," Future Generation Computer Systems, vol. 102, pp. 902–911, Jan. 2020.

[102] Y.-J. Chen, L.-C. Wang, and S. Wang, "Stochastic blockchain for iot data integrity," IEEE Transactions on Network Science and Engineering, vol. 7, pp. 373–384, Jan.-March 2020.

[103] Q. Feng, D. He, S. Zeadally, M. K. Khan, and N. Kumar, "A survey on privacy protection in blockchain system," Journal of Network and Computer Applications, vol. 126, pp. 45–58, Jan. 2019.

[104] X. Xu, Q. Liu, X. Zhang, J. Zhang, L. Qi, and W. Dou, "A blockchain-powered crowdsourcing method with privacy preservation in mobile environment," IEEE Transactions on Computational Social Systems, vol. 6, pp. 1407–1419, Dec. 2019.

[105] Y. Wu, H.-N. Dai, H. Wang, and K.-K. R. Choo, "Blockchain-based privacy preservation for 5g-enabled drone communications," IEEE Network, vol. 35, pp. 50–56, Jan./Feb. 2021.

[106] S. Dhelim, T. Kechadi, L. Chen, N. Aung, H. Ning, and L. Atzori, "Edge-enabled metaverse: The convergence of metaverse and mobile edge computing," arXiv preprint arXiv:2205.02764, 2022.

[107] J. D. N. Dionisio, W. G. B. III, and R. Gilbert, "3d virtual worlds and the metaverse: Current status and future possibilities," ACM Comput. Surv., vol. 45, jul 2013.

[108] Credit Suisse, "Metaverse: A guide to the next-gen internet," 2022. Accessed on 27.08.2022.

[109] A. Bujari, O. Gaggi, M. Luglio, C. E. Palazzi, G. Quadrio, C. Roseti, and F. Zampognaro, "Addressing the bandwidth demand of immersive applications through nfv in a 5g network," Mobile Networks and Applications, vol. 25, no. 3, pp. 1114–1121, 2020.

[110] P. Hu, S. Dhelim, H. Ning, and T. Qiu, "Survey on fog computing: architecture, key technologies, applications and open issues," Journal of Network and Computer Applications, vol. 98, pp. 27–42, 2017.

[111] A. Naouri, H. Wu, N. A. Nouri, S. Dhelim, and H. Ning, "A novel framework for mobile-edge computing by optimizing task offloading," IEEE Internet of Things Journal, vol. 8, no. 16, pp. 13065–13076, 2021.

[112] W. Zhang, J. Chen, Y. Zhang, and D. Raychaudhuri, "Towards efficient edge cloud augmentation for virtual reality mmogs," in Proceedings of the Second ACM/IEEE Symposium on Edge Computing, SEC '17, (New York, NY, USA), Association for Computing Machinery, 2017.

[113] W. Y. B. Lim, Z. Xiong, D. Niyato, X. Cao, C. Miao, S. Sun, and Q. Yang, "Realizing the Metaverse with Edge Intelligence: A Match Made in Heaven," arXiv e-prints, p. arXiv:2201.01634, Jan. 2022.

[114] K. B. Letaief, Y. Shi, J. Lu, and J. Lu, "Edge artificial intelligence for 6g: Vision, enabling technologies, and applications," IEEE Journal on Selected Areas in Communications, vol. 40, no. 1, pp. 5–36, 2022.

[115] Z. Wang, J. Qiu, Y. Zhou, Y. Shi, L. Fu, W. Chen, and K. B. Letaief, "Federated learning via intelligent reflecting surface," IEEE Transactions on Wireless Communications, vol. 21, no. 2, pp. 808–822, 2022.

[116] M. Xu, D. Niyato, J. Kang, Z. Xiong, C. Miao, and D. I. Kim, "Wireless edge-empowered metaverse: A learning-based incentive mechanism for virtual reality," in ICC 2022-IEEE International Conference on Communications, pp. 5220–5225, IEEE, 2022.

[117] Y. Sun, Z. Chen, M. Tao, and H. Liu, "Communications, caching, and computing for mobile virtual reality: Modeling and tradeoff," IEEE Transactions on Communications, vol. 67, no. 11, pp. 7573–7586, 2019.

[118] Y. Jiang, J. Kang, D. Niyato, X. Ge, Z. Xiong, C. Miao, Xuemin, and Shen, "Reliable Distributed Computing for Metaverse: A Hierarchical Game-Theoretic Approach," arXiv e-prints, p. arXiv:2111.10548, Nov. 2021.

[119] X. Huang, K. Zhang, F. Wu, and S. Leng, "Collaborative machine learning for energy-efficient edge networks in 6g," IEEE Network, vol. 35, no. 6, pp. 12–19, 2021.

[120] X. Wang, Y. Han, C. Wang, Q. Zhao, X. Chen, and M. Chen, "In-edge ai: Intelligentizing mobile edge computing, caching and communication by federated learning," IEEE Network, vol. 33, no. 5, pp. 156–165, 2019.

[121] W. Y. B. Lim, N. C. Luong, D. T. Hoang, Y. Jiao, Y.-C. Liang, Q. Yang, D. Niyato, and C. Miao, "Federated learning in mobile edge networks: A comprehensive survey," IEEE Communications Surveys & Tutorials, vol. 22, no. 3, pp. 2031–2063, 2020.

[122] X. Wang, Y. Han, V. C. M. Leung, D. Niyato, X. Yan, and X. Chen, "Convergence of edge computing and deep learning: A comprehensive survey," IEEE Communications Surveys & Tutorials, vol. 22, no. 2, pp. 869–904, 2020.

[123] Y. Liu, Y. Zhu, and J. J. Yu, "Resource-constrained federated edge learning with heterogeneous data: Formulation and analysis," IEEE Transactions on Network Science and Engineering, vol. 9, no. 5, pp. 3166–3178, 2022.

[124] T. Mikolov, A. Deoras, D. Povey, L. Burget, and J. Černocký, "Strategies for training large scale neural network language models," in 2011 IEEE Workshop on Automatic Speech Recognition & Understanding, pp. 196–201, 2011.

[125] Z. Li, S. Zhuang, S. Guo, D. Zhuo, H. Zhang, D. Song, and I. Stoica, "Terapipe: Token-level pipeline parallelism for training large-scale language models," in International Conference on Machine Learning, pp. 6543–6552, PMLR, 2021.

[126] Y. Lu, X. Huang, K. Zhang, S. Maharjan, and Y. Zhang, "Low-latency federated learning and blockchain for edge association in digital twin empowered 6g networks," IEEE Transactions on Industrial Informatics, vol. 17, no. 7, pp. 5098–5107, 2021.

[127] N.-N. Dao, Q.-V. Pham, N. H. Tu, T. T. Thanh, V. N. Q. Bao, D. S. Lakew, and S. Cho, "Survey on aerial radio access networks: Toward a comprehensive 6g access infrastructure," IEEE Communications Surveys & Tutorials, vol. 23, no. 2, pp. 1193–1225, 2021.

[128] "Europe urged to act now to build open ran ecosystem." Accessed on 24.08.2022.

[129] "Openran begins work on ai/ml." Accessed on 24.08.2022.

[130] A. S. Abdalla, P. S. Upadhyaya, V. K. Shah, and V. Marojevic, "Toward next generation open radio access networks–what o-ran can and cannot do!," IEEE Network, 2022.

[131] F. Tang, X. Chen, M. Zhao, and N. Kato, "The roadmap of communication and networking in 6g for the metaverse," IEEE Wireless Communications, pp. 1–15, 2022.

[132] D. Linthicum, "Cloud computing and the metaverse," 2022. Accessed on 27.07.2022.

[133] Y. Cai, J. Llorca, A. M. Tulino, and A. F. Molisch, "Compute-and data-intensive networks: The key to the metaverse," arXiv preprint arXiv:2204.02001, 2022.

[134] Google, "Google Stadia." Accessed on 27.08.2022.

[135] Nvidia, "Nvidia GeForce Now." Accessed on 25.08.2022.

[136] "Global IoT Market will Grow to 24.1 Billion Devices in 2030, Generating $1.5 Trillion Annual Revenue." Transforma Insights research, May, 2020. [Accessed on 09.06.2022].

[137] C. De Alwis, A. Kalla, Q.-V. Pham, P. Kumar, K. Dev, W.-J. Hwang, and M. Liyanage, "Survey on 6g frontiers: Trends, applications, requirements, technologies and future research," IEEE Open Journal of the Communications Society, vol. 2, pp. 836–886, 2021.

[138] N. Janbi, I. Katib, A. Albeshri, and R. Mehmood, "Distributed artificial intelligence-as-a-service (daiaas) for smarter ioe and 6g environments," Sensors, vol. 20, no. 20, p. 5796, 2020.

[139] S. K. Jagatheesaperumal, K. Ahmad, A. Al-Fuqaha, and J. Qadir, "Advancing education through extended reality and internet of everything enabled metaverses: Applications, challenges, and open issues," arXiv preprint arXiv:2207.01512, 2022.

[140] S. Nayak and R. Patgiri, "6g communication: Envisioning the key issues and challenges," arXiv preprint arXiv:2004.04024, 2020.

[141] M. Maier, A. Ebrahimzadeh, S. Rostami, and A. Beniiche, "The internet of no things: Making the internet disappear and" see the invisible"," IEEE Communications Magazine, vol. 58, no. 11, pp. 76–82, 2020.

[142] J. Liu, H. Liu, Y. Chen, Y. Wang, and C. Wang, "Wireless sensing for human activity: A survey," IEEE Communications Surveys & Tutorials, vol. 22, no. 3, pp. 1629–1645, 2019.

[143] Y. Cui, F. Liu, X. Jing, and J. Mu, "Integrating sensing and communications for ubiquitous iot: Applications, trends, and challenges," IEEE Network, vol. 35, no. 5, pp. 158–167, 2021.

[144] Y. Siriwardhana, P. Porambage, M. Liyanage, and M. Ylianttila, "A survey on mobile augmented reality with 5g mobile edge computing: architectures, applications, and technical aspects," IEEE Communications Surveys & Tutorials, vol. 23, no. 2, pp. 1160–1192, 2021.

[145] N. Xi, J. Chen, F. Gama, M. Riar, and J. Hamari, "The challenges of entering the metaverse: An experiment on the effect of extended reality on workload," Information Systems Frontiers, pp. 1–22, 2022.

[146] A. Plechatá, G. Makransky, and R. Böhm, "Can extended reality in the metaverse revolutionise health communication?," NPJ digital medicine, vol. 5, no. 1, pp. 1–4, 2022.

[147] Q. Yu, M. Wang, H. Zhou, J. Ni, J. Chen, and S. Céspedes, "Guest editorial special issue on cybertwin-driven 6g: Architectures, methods, and applications," IEEE Internet of Things Journal, vol. 8, no. 22, pp. 16191–16194, 2021.

[148] Q. Yu, J. Ren, Y. Fu, Y. Li, and W. Zhang, "Cybertwin: An origin of next generation network architecture," IEEE Wireless Communications, vol. 26, no. 6, pp. 111–117, 2019.

[149] F. Tang, X. Chen, T. K. Rodrigues, M. Zhao, and N. Kato, "Survey on digital twin edge networks (diten) towards 6g," IEEE Open Journal of the Communications Society, 2022.

[150] W. Saad, M. Bennis, and M. Chen, "A vision of 6g wireless systems: Applications, trends, technologies, and open research problems," IEEE network, vol. 34, no. 3, pp. 134–142, 2019.

[151] M. Z. Chowdhury, M. Shahjalal, S. Ahmed, and Y. M. Jang, "6g wireless communication systems: Applications, requirements, technologies, challenges, and research directions," IEEE Open Journal of the Communications Society, vol. 1, pp. 957–975, 2020.

[152] "Meta." Accessed on 27.07.2022.

[153] "Vr hive." Accessed on 27.07.2022.

[154] "6g-life." Accessed on 27.07.2022.

[155] "Decentralland." Accessed on 27.07.2022.

[156] "Sandbox." Accessed on 27.07.2022.

[157] "Axie infinity." Accessed on 27.07.2022.

[158] "Illuvium." Accessed on 27.07.2022.

[159] "Luxembourg metaverse." Accessed on 27.07.2022.

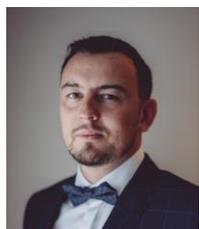
BARTLOMIEJ SINIARSKI is currently a post doctoral researcher and a project manager for the EU H2020 SPATIAL project at University College Dublin. He completed his undergraduate studies in Computer Science at University College Dublin (Ireland) and University of New South Wales (Australia). He was awarded with a doctoral degree in 2018. He has a particular interest and experience in the design of the IoT networks and in particular collecting, storing and analysing data gathered from intelligent sensors. Furthermore, he was actively involved in MSCA-ITN-ETN, ICT-52-2020 and H2020-SU-DS-2020 projects which are focused on solving problems in the area of network security, performance and management in 5G and B5G networks.

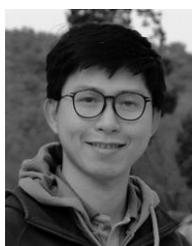
THIEN HUYNH-THE received the B.S. degree in Electronics and Telecommunication Engineering and the M.Sc. degree in Electronics Engineering from Ho Chi Minh City University of Technology and Education, Vietnam, in 2011 and 2013, respectively, and the Ph.D. degree in Computer Science and Engineering from Kyung Hee University (KHU), South Korea, in 2018. He was a recipient of the Superior Thesis Prize awarded by KHU. From March 2018 to August 2018, he was a Postdoctoral Researcher with Ubiquitous Computing Laboratory, KHU. From September 2018 to May 2022, he was a Postdoctoral Researcher with the ICT Convergence Research Center, Kumoh National Institute of Technology, South Korea. He is currently a Lecturer in Department of Computer and Communication Engineering, Ho Chi Minh City University of Technology and Education (HCMUTE), Vietnam. He was a recipient of Golden Globe Award 2020 for Vietnamese Young Scientist by Central Ho Chi Minh Communist Youth Union associated with Ministry of Science and Technology. His current research interests include digital image processing, radio signal processing, computer vision, wireless communications, IoT applications, machine learning, and deep learning.

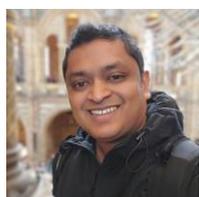
CHAMITHA DE ALWIS (Senior Member, IEEE) is a Lecturer, Researcher and Consultant in Cybersecurity. Presently he works as a Lecturer in Cybersecurity in the School of Computer Science and Technology, University of Bedfordshire, United Kingdom. He is the founder Head of the Department of Electrical and Electronic Engineering, University of Sri Jayewardenepura, Sri Lanka, where he also served as a Senior Lecturer. He received the B.Sc. (First Class Hons.) in Electronic and Telecommunication Engineering from University of Moratuwa, Sri Lanka, in 2009, and the Ph.D. in Electronic Engineering from University of Surrey, United Kingdom, in 2014. He has over 13 years of experience in the academia and the industry. He has published over 30 research articles and serves as guest editor/reviewer/TPC member for reputed journals and conferences. He was awarded several competitive research grants and actively contributes to various research projects related to network security, 5G/6G, and blockchain. He also provides consultancy services for ICT and cybersecurity related projects and activities.

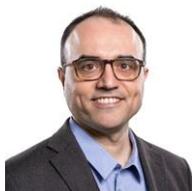
GÜRKAN GÜR (Senior Member, IEEE) is a senior lecturer at Zurich University of Applied Sciences (ZHAW) – Institute of Applied Information Technology (InIT) in Winterthur, Switzerland. He received his B.S. degree in electrical engineering in 2001 and Ph.D. degree in computer engineering in 2013 from Bogazici University in Istanbul, Turkey. His research interests include Future Internet, 5G and Beyond networks, information security, and information-centric networking. He has two patents (one in US, one in TR) and published more than 80 academic works. Currently, he is involved in EU H2020 RIA – INSPIRE-5Gplus project. He is a senior member of IEEE and a member of ACM. His research interests include Future Internet, information security, next-generation wireless networks and ICN.

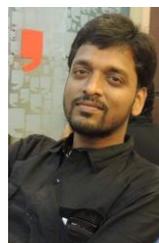
THIPPA REDDY GADEKALLU is currently working as an Associate Professor in the School of Information Technology and Engineering, Vellore Institute of Technology, Vellore, Tamil Nadu, India. He obtained his Bachelors in Computer Science and Engineering from Nagarjuna University, India, in the year 2003, Masters in Computer Science and Engineering from Anna University, Chennai, Tamil Nadu, India in the year 2011 and his Ph.D in Vellore Institute of Technology, Vellore, Tamil Nadu, India in the year 2017. He has more than 14 years of experience in teaching. He has more than 150 international/national publications in reputed journals and conferences. Currently, his areas of research include Machine Learning, Internet of Things, Deep Neural Networks, Blockchain, Computer Vision. He is an editor in several publishers like Springer, Hindawi, Plosone, Scientific Reports (Nature), Wiley. He also acted as a guest editor in several reputed publishers like IEEE, Elsevier, Springer, Hindawi, MDPI. He is recently recognized as one among the top 2% scientists in the world as per the survey conducted by Elsevier in the year 2021, 2022.

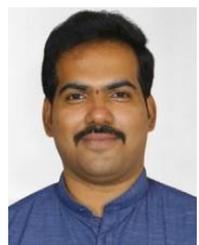
GOKUL YENDURI received his Master's degree (M.Tech., IT) from Vellore Institute of Technology in the year 2013. Currently, he is a senior research fellow at the DIVERSASIA project, co-funded by the Erasmus+ programme of the European Union. His areas of interest are machine learning and predictive analysis, software engineering, assistive technologies, and the metaverse. He has worked as an assistant professor in the past. He attended several national and international conferences, workshops, and guest lectures and published papers in peer-reviewed international journals. He is also acting as a reviewer for many prestigious peer-reviewed international journals.

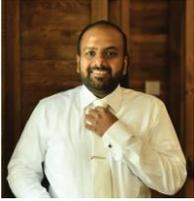

MADHUSANKA LIYANAGE (Senior Member, IEEE) is an Assistant Professor/Ad Astra Fellow and Director of Graduate Research at the School of Computer Science, University College Dublin, Ireland. He is also acting as a Docent/Adjunct Professor at the Center for Wireless Communications, University of Oulu, Finland, and Honorary Adjunct Professor at the Department of Electrical and Information Engineering, University of Ruhuna, Sri Lanka. He received his Doctor of Technology degree in communication engineering from the University of Oulu, Oulu, Finland, in 2016. From 2011 to 2012, he worked as a Research Scientist at the I3S Laboratory and Inria, Sophia Antipolis, France. He was also a recipient of the prestigious Marie Skłodowska-Curie Actions Individual Fellowship and Government of Ireland Postdoctoral Fellowship during 2018-2020. During 2015-2018, he has been a Visiting Research Fellow at the CSIRO, Australia, the Infolabs21, Lancaster University, U.K., Computer Science and Engineering, The University of New South Wales, Australia, School of IT, University of Sydney, Australia, LIP6, Sorbonne University, France and Computer Science and Engineering, The University of Oxford, U.K. He is also a senior member of IEEE. In 2020, he received the "2020 IEEE ComSoc Outstanding Young Researcher" award by IEEE ComSoc EMEA. In 2021, he was ranked among the World's Top 2% Scientists (2020) in the List prepared by Elsevier BV, Stanford University, USA. Also, he was awarded an Irish Research Council (IRC) Research Ally Prize as part of the IRC Researcher of the Year 2021 awards for the positive impact he has made as a supervisor. Dr. Liyanage's research interests are 5G/6G, SDN, IoT, Blockchain, MEC, mobile, and virtual network security. More info: www.madhusanka.com

· · ·